\documentclass[aps,prl,reprint,superscriptaddress,amsmath,amssymb,floatfix,raggedbottom]{revtex4-2}

\usepackage{graphicx}
\usepackage{tikz}
\usepackage{bm}
\usepackage{booktabs}
\usepackage{xcolor}

\usepackage{xeCJK}
\setCJKsansfont{FandolHei-Regular.otf}[BoldFont=FandolHei-Bold.otf]
\setCJKmonofont{FandolFang-Regular.otf}

\newif\ifshownotes
\shownotesfalse %

\usepackage[unicode,hidelinks]{hyperref}
\AtBeginDvi{}
\newcommand{\figtabref}[1]{\hyperref[#1]{\textcolor{red}{\ref*{#1}}}}
\newcommand{\smref}[1]{\hyperref[#1]{Sec.~\ref*{#1} of the Supplemental Material}}
\graphicspath{{figures/}}
\begin{document}

\title{Anyon Proliferation at Self-Triality: From a Lattice Model to Adjoint \texorpdfstring{QCD$_3$}{QCD3}}
\author{Yuxuan GUO  (郭雨轩)}
\email{yuxguo2024@g.ecc.u-tokyo.ac.jp}
\affiliation{Department of Physics, University of Tokyo, Japan}

\begin{abstract}
We study how competitive proliferation among three types of anyons in a self-trial $\mathbb Z_2\times\mathbb Z_2$ gauge theory can drive symmetry breaking while preserving part of the topological order.
We construct a deformation of the honeycomb color code, a spin model rooted in quantum error correction, with an $S_3$ symmetry that permutes the competing anyons, together with time-reversal and fermionic $\mathbb Z_2^{(1)}$ one-form symmetries.
We propose two dual continuum descriptions of this transition.
The bosonic description is a $\mathrm{PSU}(3)_3\times\mathrm{Spin}(8)_{-1}$ Chern--Simons--Higgs theory with a real Higgs field in the $(\mathbf8_{\mathrm{adj}},\mathbf8_v)$ representation.
Fermion-boson duality gives a particularly simple fermionic dual: $\mathrm{PSU}(3)_{3/2}$ QCD$_3$ with a single adjoint Majorana fermion and gauged fermion parity.
Both descriptions match the lattice model's symmetries and proposed phase diagram, connecting double-toric-code order to a symmetry-broken phase with residual toric-code order.
The transition is described by Higgs condensation and by a Majorana mass changing sign, respectively.
This connection to adjoint QCD$_3$ points to a possible continuous $S_3$-breaking transition from competing anyons, in contrast to the conventional first-order three-state Potts transition from competing local order parameters.
\end{abstract}
\maketitle
\textit{Introduction.---}
Topological order is a central concept in modern condensed matter physics~\cite{Wen2017Zoo,Wen2019Entanglement}.
In $(2+1)$ dimensions, it supports anyonic excitations whose fusion and braiding are encoded by unitary modular tensor categories~\cite{Kitaev2006}.
Transitions between topological phases arise in fractional quantum Hall systems~\cite{JainKivelsonTrivedi1990,Shahar1995,Shahar1996,PuSreejithJain2022,Madathil2023,ShiSenthilDoping2025,Voinea2026,ShiSenthilAnyonFluid2026,AlamKarchLuSpieler2026} and discrete gauge theories~\cite{Wegner1971,FradkinShenker1979,Trebst2007,HammaLidar2008,CastelnovoChamon2008}.
Anyon condensation relates these phases through condensable Frobenius algebras~\cite{Kong2014,Kong2021}, which specify compatible bosonic condensates and determine the resulting anyon content~\cite{BaisSchroers2002,BaisSlingerland2009,Eliens2014,Neupert2016}.
This algebraic framework does not by itself determine transition dynamics~\cite{Burnell2018,GarciaSepulvedaLu2026}.
Field-theoretic descriptions often address these dynamics by coupling fractionalized degrees of freedom to emergent gauge fields. Examples range from Chern--Simons--Higgs (CSH) theories of quantum Hall phases and related transitions~\cite{ZhangHanssonKivelson1989,Read1989,Kivelson1992,WenWu1993,ChenFisherWu1993,GoldmanSohalFradkin2019,ShiZhangSenthil2025,ShiSenthilMelting2025,CordovaGarciaSepulvedaOhmori2026,ZhouWangHe2026,LiuZhangSong2026} to recent systematic approaches to anyon condensation and proliferation~\cite{ChengSeiberg2026,ChengSeibergPeeling2026,SchaferNamekiZhengAntinucci2026,Cheng2026,GaiSchaferNameki2026,EckChen2026}.

\begin{figure*}[t]
\centering
\includegraphics[width=\textwidth]{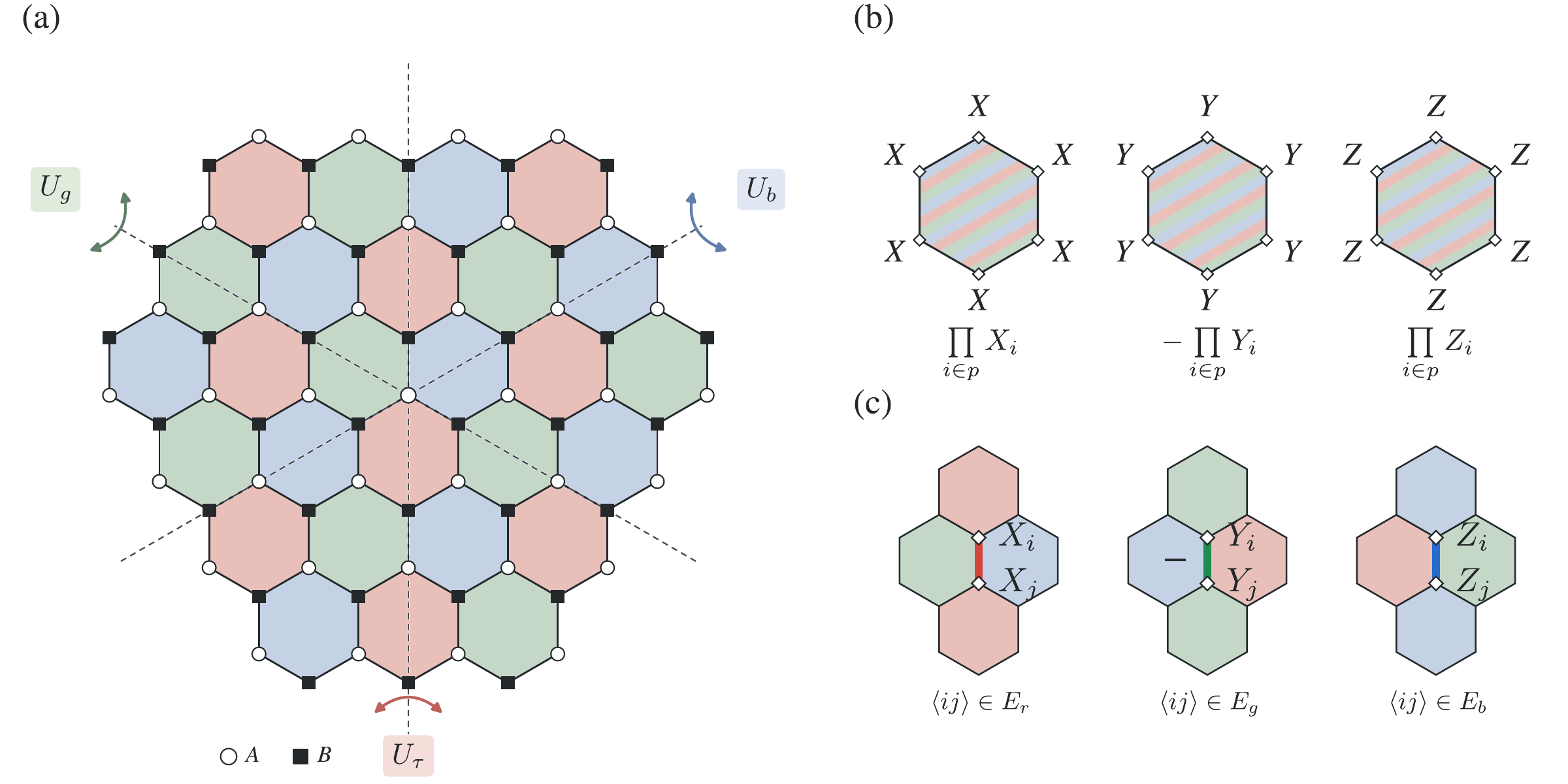}
\caption{Honeycomb realization of competing condensation channels.
(a) Qubits occupy the $A$ (open circles) and $B$ (filled squares) sublattices. The green, blue, and red (vertical) mirrors $R_g$, $R_b$, and $R_{\mathrm v}$ combine with Clifford rotations to give $U_g=R_g\prod_i\frac{X_i+Z_i}{\sqrt2}$, $U_b=R_b\prod_{i\in A}\frac{-X_i-Y_i}{\sqrt2}\prod_{j\in B}\frac{-X_j+Y_j}{\sqrt2}$, and $U_\tau=R_{\mathrm v}\prod_{i\in A}\frac{Y_i-Z_i}{\sqrt2}\prod_{j\in B}\frac{-Y_j-Z_j}{\sqrt2}$. The single-qubit Clifford factors on $A$ are $H$, $F$, and $HFH$, respectively.
(b) The three signed plaquette operators entering $H_{\mathrm{CC}}$ act on every face color, as indicated by the stripes.
(c) The signed bond operators are $X_iX_j$, $-Y_iY_j$, and $Z_iZ_j$ on $E_r$, $E_g$, and $E_b$, respectively. Edge color is determined by the adjacent face colors, independently of its direction.
The finite patch illustrates the bulk geometry.}
\label{fig:color-code-lattice}
\end{figure*}

A further challenge arises when symmetry places mutually nonlocal anyons on equal footing.
A paradigmatic example is the Fradkin--Shenker model~\cite{Tupitsyn2010}
\begin{equation}
 H_{\mathrm{FS}}=H_{\mathrm{TC}}-h_e\sum_i Z_i-h_m\sum_i X_i,
 \label{eq:fradkin-shenker}
\end{equation}
where $H_{\mathrm{TC}}$ is the toric-code Hamiltonian~\cite{Kitaev2003}, and the $h_e$ and $h_m$ terms lower the gaps of the $e$ and $m$ excitations, respectively~\cite{Vidal2009}.
On the self-dual line $h_e=h_m=h$, the $\mathbb{Z}_2^{\mathrm{em}}$ self-duality exchanges $e$ and $m$.
Numerical studies support a continuous transition, as $h$ increases, from TC topological order to an intermediate, topologically trivial phase that spontaneously breaks self-duality with twofold ground-state degeneracy~\cite{Wu2012,Somoza2021,Oppenheim2024,XuPollmannKnap2025}, followed by a symmetry-restoring Ising transition at larger $h$.
The intermediate phase reflects statistical frustration between $e$ and $m$: their mutual-semionic statistics exclude a common bosonic condensate.
Each vacuum selects one condensation channel.
Despite progress in earlier analyses and proposals~\cite{Somoza2021,Bonati2022,Oppenheim2024,Serna2024,ShiChatterjee2025,XuPollmannKnap2025}, capturing both mutual statistics and the full symmetry structure in a continuum theory remained challenging.
A recent $\mathrm{SO}(4)_{2,-2}$ CSH theory with a vector Higgs field addresses this difficulty, reproducing the phase diagram and symmetries through a Higgs transition~\cite{Ji2026}.
Related developments include a Higgs--Yukawa QED$_3$ theory~\cite{Jian2017} for self-dual multicriticality in a $U(1)_e\times U(1)_m$-enriched Fradkin--Shenker model~\cite{Dumitrescu2026} and $\mathrm{SO}(4)_{4,-4}$ CSH theories for self-dual transitions in non-Abelian $S_3$ gauge theory~\cite{LuWangVishwanath}.

In this Letter, we turn to gauge theories with self-triality~\cite{TongTurner2020,LuSun2023,LuSunYou2024,LuSunZhang2025}, where a global non-Abelian $S_3$ symmetry permutes three competing anyons, and ask how their proliferation can drive symmetry breaking and give rise to new critical behavior.
A minimal example is the $\mathbb{Z}_2\times\mathbb{Z}_2$ gauge theory whose deconfined phase has double-toric-code order $\mathrm{TC}\boxtimes\mathrm{TC}$.
The competing bosons are $e_1$, $f_1f_2$, and $m_1$.
We tune these three condensation channels with equal-strength interactions that preserve $S_3$.
Condensing any one of these bosons leaves TC topological order~\cite{Kesselring2024}, but their pairwise mutual-semionic statistics exclude a common bosonic condensate.
By analogy with the self-dual Fradkin--Shenker model, we conjecture a transition to a phase that selects one channel and breaks $S_3$ to the $\mathbb{Z}_2$ subgroup that fixes it.
This phase has three symmetry-related vacua, each retaining TC topological order:
\[
  \mathrm{TC}\boxtimes\mathrm{TC}
  \longrightarrow \bigoplus_{i=1}^{3}\mathrm{TC}_i.
\]

To study this transition, we add symmetric bond interactions to the honeycomb color code, a spin model originating in quantum error correction.
The model preserves an $S_3$ triality symmetry, time reversal, and an exact fermionic $\mathbb Z_2^{(1)}$ symmetry.
Gauging triality and then gauging back organizes the three channels into a single real Higgs multiplet, yielding a bosonic CSH description with gauge group $\mathrm{PSU}(3)\times\mathrm{Spin}(8)$.
The lattice's fermionic one-form symmetry also permits a description with gauged fermion parity $(-1)^F$~\cite{BhardwajGaiottoKapustin2017}.
Fermion-boson duality of the CSH theory provides a particularly simple realization: $\mathrm{PSU}(3)$ QCD$_3$ with a single adjoint Majorana fermion $\chi$~\cite{AharonyOrthogonal2017,ChenZimet2018}.
Both descriptions reproduce the proposed gapped phases and share a unitary $S_3\times\mathbb Z_2^{(1)}$ symmetry, while $\mathcal{CT}$ is proposed to emerge at criticality.
Intriguingly, a single mass deformation in QCD$_3$ captures this symmetry-breaking transition beyond a Landau--Ginzburg description based solely on local order parameters, reminiscent of QED$_3$ descriptions of deconfined quantum critical point (DQCP)~\cite{Senthil2004DQCP,Wang2017DQCP,Senthil2019DualityReview}.

\textit{Lattice model and microscopic symmetries.---}
To make the symmetry manifest, we construct our lattice model from the color code, familiar from quantum error correction~\cite{Bombin2006,Bombin2012,Kubica2015,Kesselring2018}.
We place one qubit on each vertex of a three-colorable honeycomb lattice and use the color-code Hamiltonian [Fig.~\figtabref{fig:color-code-lattice}(a),Fig.~\figtabref{fig:color-code-lattice}(b)],
\begin{equation}
 H_{\mathrm{CC}}=-J\sum_{p\in\mathcal P}\Bigl(\prod_{i\in p}X_i-\prod_{i\in p}Y_i+\prod_{i\in p}Z_i\Bigr),
 \label{eq:color-code-hamiltonian}
\end{equation}
where $\mathcal P$ is the set of hexagonal plaquettes and $i\in p$ labels their vertices.
The nine nontrivial bosons are labeled by a face color $c\in\{r,g,b\}$ and a Pauli index $\alpha\in\{x,y,z\}$.
Table~\figtabref{tab:color-code-anyons} gives their $\mathrm{TC}\boxtimes\mathrm{TC}$ labels. Independent row and column permutations, together with transposition, generate the anyon-permuting group $(S_3\times S_3)\rtimes\mathbb Z_2$~\cite{Kesselring2018}.

Color each edge by the color absent from its two adjacent plaquettes, and let $E_c$ denote the corresponding edge set.
The signed bond operators $X_iX_j$ on $E_r$, $-Y_iY_j$ on $E_g$, and $Z_iZ_j$ on $E_b$ are short ribbon segments for the three diagonal bosonic channels [Fig.~\figtabref{fig:color-code-lattice}(c)].
Turning on the bond interactions in any one of these channels can drive anyon condensation from color-code to TC topological order~\cite{Kesselring2024,RahmaniHaghighi2025}.
Combining the three channels with equal strength gives the self-trial model
\begin{equation}
 H_{\mathrm{STC}}=H_{\mathrm{CC}}-\lambda\Bigl(\sum_{E_r}X_iX_j-\sum_{E_g}Y_iY_j+\sum_{E_b}Z_iZ_j\Bigr).
\label{eq:hamiltonian}
\end{equation}
Here each sum runs over bonds $\langle ij\rangle$ in the indicated edge set.
The single coupling $\lambda$ tunes all three channels while preserving $S_3$.

The diagonal $S_3$ is generated by $U_\rho$ and $U_\tau$, which induce the anyon permutations $\rho=(r\,g\,b)(x\,y\,z)$ and $\tau=(g\,b)(y\,z)$ in Table~\figtabref{tab:color-code-anyons}.
Microscopically, $U_\tau$ combines reflection about the vertical axis with the Clifford rotations specified in Fig.~\figtabref{fig:color-code-lattice}(a). The cycle can be chosen as $U_\rho=U_bU_\tau$, the product of two such combined reflections (see \smref{sec:sm-lattice-symmetries} for the microscopic verification).
All interactions contain an even number of spins and preserve $\mathrm T=(\prod_i\mathrm iY_i)K$, where $K$ denotes complex conjugation in the $Z$ basis. For an even number of spins, $\mathrm T^2=1$ and $\mathrm T$ commutes with $S_3$, giving $S_3\times\mathbb Z_2^{\mathrm T}$.

\begin{table}[t]
\centering
\includegraphics[width=0.9\columnwidth]{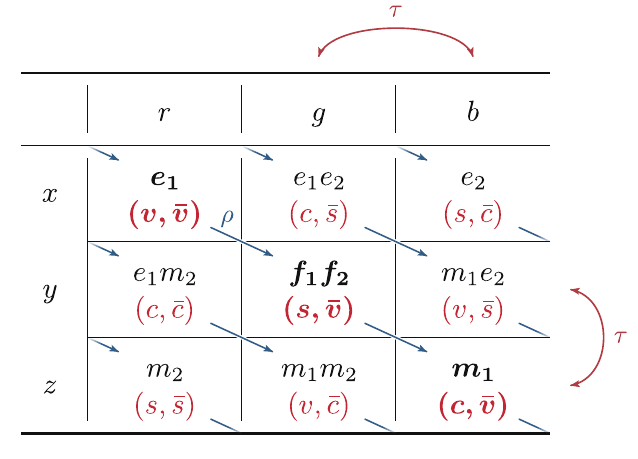}
\caption{Color-code bosons in the double-TC basis (black) and the $\mathrm{Spin}(8)_1\boxtimes\mathrm{Spin}(8)_{-1}$ basis (red). A bar denotes the negative-level factor; bold entries mark the competing condensation channels.
Blue arrows indicate the anyon permutation $\rho$ induced by $U_\rho$, with periodic continuation across table boundaries.
Red curved arrows indicate $\tau$ induced by $U_\tau$, exchanging the second and third rows and columns.}
\label{tab:color-code-anyons}
\end{table}

As $\lambda/J$ increases, we expect a transition at $\lambda=\lambda_c$ into an ordered phase that selects one condensation channel, breaking $S_3$ to $\mathbb Z_2$ and producing an imbalance among the three signed bond correlations.
An order parameter for this imbalance is
\begin{equation}
 M=X_iX_{j_r}-\bar\omega Y_iY_{j_g}+\omega Z_iZ_{j_b},
\label{eq:lattice-order-parameter}
\end{equation}
where $\omega=e^{-2\pi\mathrm i/3}$ and $j_c$ is the neighbor of $i$ along its color-$c$ bond; the three bonds meet at $120^\circ$.
The phase of $\langle M\rangle$ distinguishes the three ordered vacua.
Conjugation by $U_\rho$, $U_\tau$, and $\mathrm T$ sends $M$ to $\omega M$, $M^\dagger$, and $M^\dagger$, respectively. Since $\mathrm T$ is antiunitary, it also leaves $\langle M\rangle$ unchanged.
At $J/\lambda=0$, the model reduces to the Kekul{\'e}--Kitaev model~\cite{Quinn2015Kekule,JianYao2017FIQCP} and $S_3$ is restored.
\smref{sec:sm-kekule} shows how increasing $J/\lambda$ from this limit can lead to the same proposed intermediate $\bigoplus_{i=1}^{3}\mathrm{TC}_i$ phase.
Below we focus on the transition out of the color-code phase.

In addition to these zero-form symmetries, the model has a fermionic one-form symmetry: the three competing bosons fuse to the $S_3$-invariant fermion $e_1(f_1f_2)m_1=f_2$.
This fermion has trivial mutual braiding with all three competing bosons and survives whichever channel condenses.
Since $f_2^2=1$, its closed topological line generates a $\mathbb Z_2^{(1)}$ one-form symmetry, acting on other anyon lines through mutual braiding~\cite{Gaiotto2015}.
An exact lattice representative is
\begin{equation}
 W(C)=\mathcal P\prod_{\langle ij\rangle\in C}B_{ij},
 \label{eq:one-form-loop}
\end{equation}
where $\mathcal P$ orders bonds along a simple closed path $C$ on the honeycomb lattice, and $B_{ij}=X_iX_j,-Y_iY_j,Z_iZ_j$ on $E_r,E_g,E_b$, respectively.
These loops commute with $H_{\mathrm{STC}}$ (see \smref{sec:sm-lattice-symmetries}).
The generator's fermionic topological spin, $\theta_{f_2}=-1$~\cite{GaiottoKapustin2016,HsinLamSeiberg2019}, requires intrinsic topological order in every fully gapped phase of this bosonic model, even after $S_3$ breaking~\cite{Liu2024}.

\textit{Chern--Simons--Higgs description.---}
To construct a CSH description of the transition, we express the color-code topological order as $\mathrm{Spin}(8)_1\boxtimes\mathrm{Spin}(8)_{-1}$~\cite{Kesselring2018}, with opposite chiral central charges $c_-=4$ and $-4$, respectively.
The $\mathrm{Spin}(8)_1$ factor has three fermionic anyons $v$, $s$, and $c$, corresponding to the vector representation $\mathbf8_v$ and two chiral spinor representations $\mathbf8_s,\mathbf8_c$, each of dimension eight~\cite{TongTurner2020,Slansky1981}.
They obey $v^2=s^2=c^2=1$ and $v s=c$ (and cyclic permutations), with mutual braiding phase $-1$ between distinct species.
Table~\figtabref{tab:color-code-anyons} identifies the color-code bosons with fermion pairs from opposite sectors, providing the dictionary for expressing the diagonal color--Pauli $S_3$ in this basis.
Its normal $\mathbb Z_3$ subgroup is generated by ${U}_\rho$, which cycles $v\mapsto s\mapsto c\mapsto v$ and leaves the negative-level factor unchanged.
The transposition ${U}_\tau$ generates a nonnormal $\mathbb Z_2$ subgroup and exchanges $s\leftrightarrow c$ and $\bar s\leftrightarrow\bar c$, fixing $v$ and $\bar v$.

A direct description of the channels $(v,\bar v)$, $(s,\bar v)$, and $(c,\bar v)$ requires matter fields in three inequivalent representations of the positive-level $\mathrm{Spin}(8)_1$ factor, making their symmetry-preserving interactions and competing dynamics difficult to organize.
We instead gauge the normal $\mathbb Z_3$ to combine the channels into a single anyon species~\cite{Barkeshli2019}.
We choose the gauging that maps the positive-level factor $\mathrm{Spin}(8)_1$ to $\mathrm{SU}(3)_3$, leaving $\mathrm{Spin}(8)_{-1}$ unchanged; the orbit $\{v,s,c\}$ then becomes the non-Abelian adjoint anyon $\mathbf8_{\mathrm{adj}}$~\cite{Teo2015,Cui2016,CordovaGarciaSepulveda2026}.
The three channels thus combine into the boson $(\mathbf8_{\mathrm{adj}},\bar v)$, whose two factors each have topological spin $-1$.
We describe the dynamics of this channel using a single real $8\times8$ Higgs field $\Phi\in(\mathbf8_{\mathrm{adj}},\mathbf8_v)$.
Under $(g,h)\in\mathrm{SU}(3)\times\mathrm{Spin}(8)$, it transforms as $\Phi\mapsto R_{\mathrm{adj}}(g)\Phi R_v(h)^{\mathsf T}$.
To recover the original topological order while retaining this Higgs description, we gauge the dual, nonanomalous $\mathbb Z_3^{(1)}$ center symmetry, under which $\Phi$ is neutral~\cite{Tachikawa2020}.
This replaces $\mathrm{SU}(3)_3$ by $\mathrm{PSU}(3)_3\simeq\mathrm{Spin}(8)_1$ and restores the global $\mathbb Z_3$ as a magnetic symmetry.
The newly allowed monopole junctions on the adjoint line resolve the boson back into $(v,\bar v)$, $(s,\bar v)$, and $(c,\bar v)$~\cite{HsinLamSeiberg2019} (see \smref{sec:sm-triality-gauging}).
The resulting CSH theory is
\begin{equation}
\begin{aligned}
 \mathcal L_{\mathrm{CSH}}&={}
 \operatorname{Tr}\!\bigl[(D^\mu\Phi)^{\mathsf T}D_\mu\Phi\bigr]
 +r\operatorname{Tr}(\Phi^{\mathsf T}\Phi)+\mathcal V(\Phi)\\
 &-3\mathrm i\operatorname{CS}_{\mathfrak{su}(3)}[a]
 +\mathrm i\operatorname{CS}_{\mathfrak{spin}(8)}[b]+\frac1{g_a^2}\operatorname{Tr}f_a^2
 +\frac1{g_b^2}\operatorname{Tr} f_b^2
\end{aligned}
\label{eq:csh}
\end{equation}
Here $a$ and $b$ are dynamical Hermitian $\mathrm{PSU}(3)$ and $\mathrm{Spin}(8)$ connections, with $D_\mu\Phi=\partial_\mu\Phi-\mathrm i(a_{\mathrm{adj}})_\mu\Phi+\mathrm i\Phi(b_v)_\mu$. Our conventions are specified in \smref{sec:sm-qcd-regulator}.
The mass parameter $r$ tunes the transition, while $\mathcal{V}(\Phi)$ includes single- and multi-trace terms such as $\operatorname{Tr}[(\Phi^{\mathsf T}\Phi)^2]$ and $[\operatorname{Tr}(\Phi^{\mathsf T}\Phi)]^2$, and the determinant (baryon) term $\det\Phi$~\cite{Aharony2016Baryons}.
We choose the Higgs potential to favor $\langle\Phi\rangle=0$ for $r>0$ and a full-rank Higgs configuration $\langle\Phi\rangle=v_{\mathrm H}\mathbf{1}_8$ for $r<0$, with $v_{\mathrm H}$ the Higgs amplitude and $r=0$ set at the transition.

\textit{Internal symmetries of the CSH theory.---}
The cyclic part of $S_3$ is encoded in the gauge structure: $\pi_1(\mathrm{PSU}(3))=\mathbb Z_3$ allows monopoles with magnetic charge $q\in\mathbb Z_3$.
The corresponding magnetic symmetry is generated on a closed surface $\Sigma$ by
\begin{equation}
 \mathcal{U}_\rho(\Sigma)=\exp\!\left[-\frac{2\pi\mathrm i}{3}\int_\Sigma w_2^{(3)}(a)\right],
\label{eq:magnetic-generator}
\end{equation}
where $w_2^{(3)}(a)\in H^2(X_3,\mathbb Z_3)$ is the obstruction to lifting the $\mathrm{PSU}(3)$ bundle on spacetime $X_3$ to $\mathrm{SU}(3)$~\cite{KapustinSeiberg2014}.
Gauging this magnetic symmetry imposes $w_2^{(3)}(a)=0$, recovering the $\mathrm{SU}(3)$ theory.
The magnetic symmetry cyclically permutes the resolved lines, identifying $\mathbb Z_3^{\mathcal M}$ with the microscopic cycle.
We identify the lattice $M$ with a scalar, gauge-invariant dressed monopole operator $\mathcal M$ ~\cite{Borokhov2002,Assel2019} that creates one unit of positive magnetic charge, $q=1\pmod 3$, so that $\mathcal{U}_\rho\mathcal M\mathcal{U}_\rho^{-1}=\omega\mathcal M$.

The transposition of channels is implemented internally by the unitary symmetry $\mathcal{U}_\tau\equiv\mathcal C$ combining $\mathrm{PSU}(3)$ charge conjugation with the $\mathrm{Spin}(8)$ outer automorphism exchanging $\bar s$ and $\bar c$~\cite{CordovaOrthogonal2018} (see \smref{sec:sm-spin8-psu3} for its matrix realization).\footnote{The lattice $U_\rho,U_\tau$ and internal $\mathcal{U}_\rho,\mathcal{U}_\tau$ differ by spatial transformations in the proposed infrared matching; see \smref{sec:sm-qcd-ct}.}
Without $\det\Phi$, the two conjugations leave the action unchanged independently, giving an extra $\mathbb Z_2$ symmetry.
Since $\det\Phi$ changes sign under either alone, this interaction is invariant only under their combined action.
The combined operation sends $\mathcal M\mapsto\mathcal M^\dagger$ and $\mathcal{U}_\rho\mapsto\mathcal{U}_\rho^{-1}$, realizing the internal $S_3\cong\mathbb Z_3^{\mathcal M}\rtimes\mathbb Z_2^{\mathcal C}$ action on the order parameter.

The CSH theory also reproduces the lattice fermionic $\mathbb Z_2^{(1)}$ symmetry.
The central element $z_v\in\mathrm{Spin}(8)$ acts trivially on the vector representation, $R_v(z_v)=\mathbf1_8$, so the matter field does not screen the corresponding center symmetry~\cite{CordovaOrthogonal2018}.
Its generator is a line carrying $z_v$ holonomy, dressed with electric charge as required by the Chern--Simons term.
Restricting the $\mathrm{Spin}(8)_{-1}$ factor to a Cartan $U(1)$ subgroup, the $z_v$ vortex corresponds to a $\pi$ flux in a $U(1)_{-4}$ theory and hence carries topological spin $\theta=-1$, matching the fermionic statistics of the lattice symmetry generator (see \smref{sec:sm-spin8-psu3}).

\textit{Matching the gapped phases.---}
We now show how the CSH theory reproduces the two proposed phases.
For $r>0$, integrating out the massive Higgs field leaves $\mathrm{PSU}(3)_3\boxtimes\mathrm{Spin}(8)_{-1}$ at long distances.
Since $\mathrm{PSU}(3)_3\simeq\mathrm{Spin}(8)_1$, this reproduces the $\mathrm{TC}\boxtimes\mathrm{TC}$ topological order.
For $r<0$, the full-rank condensate $\langle\Phi\rangle=v_{\mathrm H}\mathbf1_8$ locks the two gauge factors through $R_{\mathrm{adj}}(g)R_v(h)^{\mathsf T}=\mathbf1_8$.
The adjoint embedding of $\mathrm{PSU}(3)$ into $\mathrm{Spin}(8)$~\cite{KomargodskiAdjoint2021} gives the unbroken gauge group $\mathrm{PSU}(3)_{\mathrm d}\times\mathbb Z_2^v$, where $\mathrm d$ labels the diagonal subgroup and $\mathbb Z_2^v=\{1,z_v\}$ (see \smref{sec:sm-spin8-psu3}).
This independent discrete gauge sector is inherited from $\mathrm{Spin}(8)$: it is invisible to the vector matter but detectable by spinor Wilson lines.
The embedding index is three, so the $\mathrm{Spin}(8)_{-1}$ term induces level $-3$ on the diagonal connection $a_{\mathrm d}$, canceling the original level $+3$ and leaving $\mathrm{PSU}(3)$ Yang--Mills theory~\cite{GarciaSepulvedaLu2026}.
The covering $\mathrm{SU}(3)$ theory confines~\cite{Teper1999}, leaving its electric $\mathbb Z_3^{(1)}$ center symmetry unbroken.
Gauging this center gives $\mathrm{PSU}(3)$ with spontaneously broken magnetic $\mathbb Z_3^{\mathcal M}$ and three vacua~\cite{Gaiotto2015}.

To determine the residual $\mathbb Z_2^v$ action, we use the $U(1)_{-4}$ Cartan restriction above.
A charge-two Higgs field retains $\mathbb Z_2^v$; dualizing it gives a Chern Simons theory $\mathcal L_K$ with the $K$ matrix~\cite{KapustinSeiberg2014,WenZee1992}
\begin{equation}
 K=\begin{pmatrix}-4&2\\2&0\end{pmatrix}
 \sim K_{\mathrm{TC}}=\begin{pmatrix}0&2\\2&0\end{pmatrix},
\label{eq:residual-k-matrix}
\end{equation}
where $\sim$ denotes a unimodular change of basis.
Thus the residual theory is the toric code rather than the double-semion theory, with its fermion carrying the inherited one-form symmetry; the derivation is given in \smref{sec:sm-higgs}.
Each magnetic vacuum therefore retains one TC sector, giving $\bigoplus_{i=1}^3\mathrm{TC}_i$ and the expected $S_3\to\mathbb Z_2$ breaking.

The antiunitary operation $\mathcal{CT}$ commutes with $\mathcal{U}_\rho$ and sends $\mathcal M$ to $\mathcal M^\dagger$~\cite{Cordova2018}, matching the action of the microscopic time reversal $\mathrm T$ on the lattice order parameter.
Both gapped phases preserve $\mathrm T$.
Together, these observations motivate an emergent $\mathcal{CT}$ symmetry at the proposed critical point.
Our results are summarized below, with equivalences understood in the infrared:
\begin{equation}
\begin{tikzpicture}[baseline=(current bounding box.center),
 every node/.style={font=\small,inner sep=1pt}]
 \node (uv) at (0,0) {$\mathcal L_{\mathrm{CSH}}$};
 \node[minimum height=0.35cm] (h) at (-1.75,-1.12) {$\mathrm{PSU}(3)$ ${\mathrm{YM}}\boxtimes\mathrm{CS}[K]
 $};
 \node[minimum height=0.35cm] (u) at (1.75,-1.12) {$
  \mathrm{PSU}(3)_3\boxtimes\mathrm{Spin}(8)_{-1}
 $};
 \node[anchor=base west,inner sep=0pt] at (u.base east) {$\,\simeq$};
 \node[anchor=base west] (hp) at ([xshift=0.24cm,yshift=-0.58cm]h.west) {$
  \simeq\textstyle\bigoplus_{i=1}^{3}\mathrm{TC}_i
 $};
 \node[anchor=base] (up) at (1.90,-1.70) {$
  \mathrm{Spin}(8)_1\boxtimes\mathrm{Spin}(8)_{-1}
  \simeq\mathrm{TC}\boxtimes\mathrm{TC}
 $};
 \draw[->] (uv.south west) -- node[above left,pos=.52] {$r<0$} (h.north);
 \draw[->] (uv.south east) -- node[above right,pos=.52] {$r>0$} (u.north);
\end{tikzpicture}
\label{eq:csh-phase-flows}
\end{equation}

\textit{QCD$_3$ description.---}
A fermion-boson duality for orthogonal gauge groups gives a simpler description: a single adjoint Majorana fermion coupled to a $\mathrm{PSU}(3)$ gauge field.
To obtain it, we first gauge the CSH theory's fermionic $\mathbb Z_2^{(1)}$ symmetry on a spin manifold~\cite{GaiottoKapustin2016,HsinLamSeiberg2019}, replacing $\mathrm{Spin}(8)_{-1}$ by $\mathrm{SO}(8)_{-1}$.
The magnetic $\mathbb Z_2$ of this orthogonal gauge theory is fermion parity.
Replacing $a$ by a background connection $A$, we apply this duality~\cite{Metlitski2017Majorana,AharonyOrthogonal2017}, $\mathrm{SO}(N)_{-1}+N\Phi+\det\Phi\longleftrightarrow N\chi$, at $N=8$~\cite{ChenZimet2018}, understood up to the background counterterm specified in \smref{sec:sm-qcd-regulator}.
These eight components form a single adjoint Majorana of $\mathrm{PSU}(3)$; promoting $A$ to the dynamical field $a$ gives
\begin{equation}
 \begin{aligned}
 \mathcal L_{\mathrm{QCD}}={}
 \operatorname{Tr}\bigl[\bar\chi\gamma^\mu
 (\partial_\mu\chi-\mathrm i[a_\mu,\chi])\bigr]
 +m\operatorname{Tr}(\bar\chi\chi)
 +\frac1{g^2}\operatorname{Tr}f_a^2,
 \end{aligned}
 \label{eq:qcd3}
\end{equation}
where $\chi=\chi^At^A$ and $t^A=\lambda_A/2$, with $\lambda_A$ the Gell--Mann matrices.
We define the single-flavor adjoint theory ($N_f=1$) using Pauli--Villars regularization~\cite{Witten2016Fermion,Seiberg2016DualityWeb,AharonyOrthogonal2017}, choosing $m>0$ as the trivial superconducting reference.
In the standard convention, the massless theory is denoted $\mathrm{PSU}(3)_{3/2}+\mathrm{adj}\,\chi$~\cite{Gomis2018}. The magnetic $\mathbb Z_3^{\mathcal M}$ symmetry and charge conjugation $\mathcal C:(a,\chi)\mapsto(-a^{\mathsf T},-\chi^{\mathsf T})$ realize the same internal $S_3$ symmetry as the CSH theory.

Under mass deformations, we expect each gapped branch to remain connected to its large-mass limit. Integrating out the fermion at fixed $a$ in this limit determines the gauge and gravitational Chern--Simons terms.
This gives the two phases below, where $[p-\mathrm i p]^{\otimes n}$ denotes an invertible spin phase formed by $n$ stacked $p-\mathrm i p$ superconductors, with $c_-^{\mathrm{inv}}=-n/2$~\cite{Seiberg2016DualityWeb}:
\begin{equation}
\begin{tikzpicture}[baseline=(current bounding box.center),
 every node/.style={font=\small,inner sep=1pt}]
 \node (uv) at (0,0) {$\mathcal L_{\mathrm{QCD}}$};
 \node (h) at (-1.90,-1.08) {$\mathrm{PSU}(3)$ ${\mathrm{YM}}\boxtimes\text{trivial SC}$};
 \node (u) at (1.90,-1.08) {$\mathrm{PSU}(3)_3\boxtimes[p-\mathrm i p]^{\otimes 8}$};
 \draw[->] (uv.south west) -- node[above left,pos=.52] {$m>0$} (h.north);
 \draw[->] (uv.south east) -- node[above right,pos=.52] {$m<0$} (u.north);
\end{tikzpicture}
\label{eq:qcd-phase-flows}
\end{equation}
For $m<0$, the eight superconducting layers contribute $c_-^{\mathrm{inv}}=-4$, canceling the gauge contribution $c_-=4$ and yielding a nonchiral phase.
Although the Pauli--Villars regulator explicitly breaks time reversal, both gapped phases preserve it.
For $m<0$, this symmetry is realized through level--rank duality relating the phase to its $\mathcal{CT}$ conjugate~\cite{HsinSeiberg2016}.
This motivates our proposal that lattice time reversal emerges at criticality as a $\mathcal{CT}$ symmetry relating infrared-dual QCD theories.
Assuming this infrared duality, the singlet mass operator $\operatorname{Tr}(\bar\chi\chi)$ is even under the emergent $\mathcal{CT}$ symmetry (see \smref{sec:sm-qcd-ct}).

Then, gauging $(-1)^F$ gives the bosonic theory $\mathcal L_{\mathrm{QCD}}/(-1)^F$.
By Kitaev's sixteen-fold way~\cite{Kitaev2006}, the trivial superconductor and eight copies of $p-\mathrm i p$ become TC and $\mathrm{Spin}(8)_{-1}$, respectively.
This reproduces both CSH phases in Eq.~\eqref{eq:csh-phase-flows}, with the fermion line generating the same fermionic $\mathbb Z_2^{(1)}$ symmetry.
Near criticality, the local bond operator $X_iX_{j_r}-Y_iY_{j_g}+Z_iZ_{j_b}$ corresponds to the singlet mass operators $\operatorname{Tr}(\Phi^{\mathsf T}\Phi)\longleftrightarrow-\operatorname{Tr}(\bar\chi\chi)$.

Finally, we comment on the continuity of the transition. For adjoint $\mathrm{SU}(N)_k$ QCD$_3$ with $k\ge N/2$, supersymmetric considerations led Gomis et al.~\cite{Gomis2018} to propose a direct transition between the two large-mass phases ($|m|\gg g^2$).
Continuity with the controlled large-$k$ regime further suggests that the transition is governed by a CFT.
Our theory follows from their $N=3$, $k=3/2$ theory by gauging the $\mathbb Z_3$ center one-form symmetry and fermion parity.
If this transition is continuous, its CFT would describe $S_3$ symmetry breaking through competition among mutually nonlocal anyons, in contrast to the conventional first-order three-state Potts transition with the same $S_3\times\mathbb Z_2^{\mathrm T}$ zero-form symmetry~\cite{BloteSwendsen1979,GavaiKarschPetersson1989,JankeVillanova1997,ChesterSu2025}.
The conformal bootstrap could further test this proposed critical behavior~\cite{Rattazzi2008Bootstrap}, potentially incorporating line defects~\cite{GaiottoMazacPaulos2014}.

\textit{Conclusion and discussion.---}
We have constructed a deformation of the honeycomb color code and proposed dual CSH and adjoint QCD$_3$ descriptions of a transition where competing anyons break $S_3$ while preserving toric-code order.
The lack of perturbative control in these continuum theories motivates numerical studies of the transition's continuity and emergent time reversal.
Possible approaches include simulations inspired by recent quantum-double studies~\cite{FloresCalderonPollmannKnap2026}, projected entangled-pair states (PEPS) for the nonchiral phases~\cite{IqbalSchuch2021,XuGarreRubioSchuch2022}, and fuzzy-sphere methods for gauge theories~\cite{Zhu2023FuzzySphere,Huffman2026QCD} and quantum Hall transitions~\cite{Voinea2026,ZhouWangHe2026,ZhouHe2026Fuzzy}.

Competition among mutually nonlocal anyons offers a route to symmetry breaking beyond a Landau--Ginzburg description based solely on local order parameters.
Generalized gauging~\cite{WangCompatible2026,ChenLiuLuTantivasadakarn2026,EckFusion2025} may extend this mechanism to competing non-Abelian bosons in symmetry-enriched topological phases~\cite{EssinHermele2013,Barkeshli2019,ChengHsinJian2023,EckHuston2026}.
A natural question is whether QCD$_3$ theories with fermions in more general tensor representations~\cite{ChoiDelmastroGomisKomargodski2020}, after suitable discrete gauging, describe such transitions.
The present spin model and its extensions could provide access to QCD$_3$ infrared dynamics without introducing microscopic lattice fermions or discretizing Chern--Simons terms, complementing direct lattice gauge constructions~\cite{ChenZimet2018,XuChen2025}.

\begin{acknowledgments}
\textit{Acknowledgments.---}
Y.G. thanks Chong Wang for clarifying the symmetry properties of the model, Youxuan Wang for discussions at an early stage of this work, and Yuto Ashida for helpful suggestions on the presentation of the manuscript.
Y.G. acknowledges financial support from the Global Science Graduate Course (GSGC) program at the University of Tokyo.
We acknowledge the use of ChatGPT 5.6 and 6.0 for assistance in preparing figures and tables and checking algebraic calculations related to anyon condensation.
\end{acknowledgments}

\bibliography{references}

\clearpage
\onecolumngrid
\setcounter{section}{0}
\setcounter{secnumdepth}{1}
\renewcommand{\thesection}{\Roman{section}}
\renewcommand{\theHsection}{supp.\arabic{section}}
\setcounter{equation}{0}
\renewcommand{\theequation}{S\arabic{equation}}
\renewcommand{\theHequation}{supp.\arabic{equation}}
\setcounter{table}{0}
\renewcommand{\thetable}{S\arabic{table}}
\renewcommand{\theHtable}{supp.\arabic{table}}
\begin{center}
\textbf{Supplemental Material for}\\[3pt]

\textbf{Anyon Proliferation at Self-Triality: From a Lattice Model to Adjoint QCD$_3$}
\end{center}
\section{Microscopic realization of the lattice symmetries}
\label{sec:sm-lattice-symmetries}

We verify the combined reflections specified in Fig.~\figtabref{fig:color-code-lattice}(a).
The mirrors preserve the $A$ and $B$ sublattices separately and permute face and edge colors as $R_g:(r\,b)$, $R_b:(r\,g)$, and $R_{\mathrm v}:(g\,b)$.
Their single-qubit Clifford factors on $A$ are $u=H,F,HFH$, respectively, with $H=(X+Z)/\sqrt2$ and $F=-(X+Y)/\sqrt2$; on $B$ they are $u^*$.
Here the star denotes entrywise conjugation in the $Z$ basis, so both factors are unitary.
Direct Pauli conjugation gives
\begin{equation}
\begin{array}{c|ccc|ccc}
 &\multicolumn{3}{c|}{A:\ u\sigma u^\dagger}
 &\multicolumn{3}{c}{B:\ u^*\sigma u^{\mathsf T}}\\
 u&X&Y&Z&X&Y&Z\\ \hline
 H&Z&-Y&X&Z&-Y&X\\
 F&Y&X&-Z&-Y&-X&-Z\\
 HFH&-X&-Z&-Y&-X&Z&Y
\end{array}.
\label{eq:sm-lattice-pauli-actions}
\end{equation}

On an $A$--$B$ bond, these conjugate Clifford factors permute $(XX,-YY,ZZ)$ without extra signs.
For example, $F$ sends $X_iX_j$ to $Y_i(-Y_j)=-Y_iY_j$ and $-Y_iY_j$ to $-X_i(-X_j)=X_iX_j$.
Combining this action with the spatial reflection gives $U_g:XX_r\leftrightarrow ZZ_b$, $U_b:XX_r\leftrightarrow-YY_g$, and $U_\tau:-YY_g\leftrightarrow ZZ_b$, with the third channel fixed in each case; the subscripts indicate the edge sets.
Thus $U_\rho=U_bU_\tau$ cycles $XX_r\to-YY_g\to ZZ_b\to XX_r$, realizing the simultaneous color and Pauli cycle in Table~\figtabref{tab:color-code-anyons}.

The same signs preserve the plaquette terms.
Each hexagon contains three sites of each sublattice, so the signed plaquette operators $(S_p^x,S_p^y,S_p^z)=(\prod_{i\in p}X_i,-\prod_{i\in p}Y_i,\prod_{i\in p}Z_i)$ undergo the same Pauli permutations at the reflected plaquette.
For example, $U_b S_p^xU_b^{-1}=(-1)^3\prod_{i\in R_b(p)}Y_i=S_{R_b(p)}^y$.
Their sum is therefore invariant after summing over plaquettes, establishing $U_\ell H_{\mathrm{STC}}U_\ell^{-1}=H_{\mathrm{STC}}$ for $\ell=g,b,\tau$.

The chosen Clifford phases satisfy $H^2=F^2=(HF)^3=1$.
Together with the geometric mirror relations, these imply $U_\rho^3=U_\tau^2=1$ and $U_\tau U_\rho U_\tau^{-1}=U_\rho^{-1}$ on the full Hilbert space, with symmetry-compatible boundary conditions.
For a periodic honeycomb with $N_A=N_B$, the total number of spins $N$ is even.
Single-spin time reversal changes the sign of every Clifford factor in Eq.~\eqref{eq:sm-lattice-pauli-actions}, so $\mathrm T U_\ell\mathrm T^{-1}=(-1)^N U_\ell=U_\ell$ and $\mathrm T^2=1$.
Since every Hamiltonian term has even spin weight, $\mathrm T$ is also a symmetry, giving the stated $S_3\times\mathbb Z_2^{\mathrm T}$ subgroup.

\paragraph*{Exact one-form symmetry.}
For a plaquette term $S_p^\alpha$, only bonds meeting $p$ at one vertex can anticommute.
They all have the color of $p$ and the same commutation sign; a closed $C$ contains an even number of them, so $[W(C),S_p^\alpha]=0$.
At each vertex of $C$, two bonds are occupied and one is unused.
For an unused $XX$ bond, the occupied $-YY$ and $ZZ$ bonds give a local factor $YZ\propto X$, commuting with $X$.
An occupied $XX$ bond instead meets an occupied $-YY$ or $ZZ$ bond at each endpoint, giving a local $Z$ or $Y$ factor that anticommutes with $X$.
The two endpoints contribute cancelling minus signs.
The other colors follow by cyclic permutation, so $[W(C),B_{ij}]=0$ and hence $[W(C),H_{\mathrm{STC}}]=0$.
The even length of $C$ makes the overall phase real, ensuring that $W(C)$ is Hermitian and $W(C)^2=1$.

\section{Kekul{\'e}--Kitaev limit and the ordered toric-code phase}
\label{sec:sm-kekule}

At $J=0$, the face-induced edge coloring gives the Kekul{\'e}--Kitaev model~\cite{Quinn2015Kekule}.
Using the standard representation $\sigma_i^\alpha=\mathrm i b_i^\alpha c_i$, with $b_i^xb_i^yb_i^zc_i=1$~\cite{Kitaev2006}, orient every bond from sublattice $A$ to $B$.
On a color-$c$ bond carrying Pauli component $\alpha$, define $\widetilde u_{ij}=s_c\mathrm i b_i^\alpha b_j^\alpha$, where $(s_r,s_g,s_b)=(1,-1,1)$.
The signed bond is then $B_{ij}=-\mathrm i\widetilde u_{ij}c_ic_j$, with static $\mathbb Z_2$ links $\widetilde u_{ij}=\pm1$.

The color-code terms become six-fermion interactions in this representation.
For a red face, let $Q_{p,g}$ and $Q_{p,b}$ be the products of its three disjoint green and blue boundary bonds, respectively.
The signed plaquettes of Sec.~\ref{sec:sm-lattice-symmetries} obey $Q_{p,g}=S_p^y$, $Q_{p,b}=S_p^z$, and $Q_{p,g}Q_{p,b}=W_p=S_p^x$, where $W_p=W(\partial p)$.
Cyclically permuting colors gives the same identity on every face.
Thus, in the flux-free sector $W_p=1$,
\begin{equation}
 H_{\mathrm{STC}}\big|_{W_p=1}
 =\mathrm i\lambda\sum_{\langle ij\rangle}\widetilde u_{ij}c_ic_j
 -2J\sum_p Q_p-JN_p,
\label{eq:sm-kekule-majorana}
\end{equation}
where $N_p$ is the number of faces and either alternating boundary matching defines the same $Q_p$ in this flux sector.
For example, label a red hexagon cyclically by $1,\ldots,6$, with $1,3,5\in A$ and green bonds $(12),(34),(56)$.
Then
\begin{equation*}
 Q_p=B_{12}B_{34}B_{56}
 =\mathrm i\widetilde u_{12}\widetilde u_{34}\widetilde u_{56}\,c_1c_2c_3c_4c_5c_6.
\end{equation*}
The link factors make this six-Majorana operator gauge invariant; in the uniform $\widetilde u_{ij}=1$ gauge it reduces to $Q_p=\mathrm i c_1c_2c_3c_4c_5c_6$, with the factor $\mathrm i$ ensuring Hermiticity.

For a bond mean-field description uniform within each color, let $\beta_c=\langle B_{ij}\rangle$ for $\langle ij\rangle\in E_c$.
These real bilinear averages decompose into the symmetric singlet $\beta_0=(\beta_r+\beta_g+\beta_b)/3$ and an $S_3$ doublet encoded by $\varphi=\beta_r+\bar\omega\beta_g+\omega\beta_b$, the mean-field counterpart of $M$.
The cyclic permutation sends $\varphi\mapsto\omega\varphi$.
At this mean-field level, the same-color three-bond interactions motivate the color-symmetric cubic combination
\begin{equation*}
 \sum_{c=r,g,b}\beta_c^3
 =3\beta_0^3+2\beta_0|\varphi|^2+\frac19(\varphi^3+\varphi^{*3}).
\end{equation*}
At fixed $\beta_0$, this gives the quadratic and cubic structures used in $V_{\mathrm K}$ below.

We now identify the ungauged low-energy matter fields in the uniform flux-free gauge $\widetilde u_{ij}=1$.
With unit bond length, choose the $A$-to-$B$ vectors $\boldsymbol\delta_1=(0,1)$ and $\boldsymbol\delta_{2,3}=(\mp\sqrt3/2,-1/2)$.
Using Fourier phases at the actual site positions and $\mathbf c_{\mathbf k}=(c_{A,\mathbf k},c_{B,\mathbf k})^{\mathsf T}$, the $J=0$ Hamiltonian is $H_0=\frac14\sum_{\mathbf k}\mathbf c_{-\mathbf k}^{\mathsf T}h_M(\mathbf k)\mathbf c_{\mathbf k}$, with
\begin{equation}
\begin{aligned}
 h_M(\mathbf k)&=2\lambda\begin{pmatrix}0&\mathrm i f(\mathbf k)\\-\mathrm i f(\mathbf k)^*&0\end{pmatrix},
 &\varepsilon_\pm(\mathbf k)&=\pm2\lambda|f(\mathbf k)|,\\
 f(\mathbf k)&=e^{\mathrm i k_y}+2e^{-\mathrm i k_y/2}\cos(\sqrt3 k_x/2).
\end{aligned}
\label{eq:sm-kekule-bands}
\end{equation}
In the two-site Brillouin zone, the bands touch at $\mathbf K=(4\pi/(3\sqrt3),0)$ and $\mathbf K'=-\mathbf K$.
Expanding $f(\mathbf K+\mathbf q)=\frac{3}{2}(-q_x+\mathrm i q_y)+O(q^2)$ gives the cone velocity $v=3\lambda$.
The Majorana condition $c_{s,\mathbf K+\mathbf q}^\dagger=c_{s,\mathbf K'-\mathbf q}$, with $s=A,B$, relates the two valleys, leaving four real low-energy components.

Near $\mathbf K$, the kinetic matrix is $v(q_x\sigma_2-q_y\sigma_1)$.
With $U=(1-\mathrm i\sigma_1)/\sqrt2$, define the two Majorana spinors by
\begin{equation}
\begin{aligned}
 \xi_1(\mathbf q)&=\frac{U\mathbf c_{\mathbf K+\mathbf q}+U^*\mathbf c_{\mathbf K'+\mathbf q}}{\sqrt2},\\
 \xi_2(\mathbf q)&=\frac{U\mathbf c_{\mathbf K+\mathbf q}-U^*\mathbf c_{\mathbf K'+\mathbf q}}{\mathrm i\sqrt2}.
\end{aligned}
\label{eq:sm-kekule-fields}
\end{equation}
They obey $\xi_{a,\alpha}(\mathbf q)^\dagger=\xi_{a,\alpha}(-\mathbf q)$, where $\alpha$ labels the rotated sublattice components.

The lattice Clifford rotations act on the auxiliary $b_i^\alpha$ Majoranas, while $c_i$ follows the spatial operation, up to a $\mathbb Z_2$ gauge transformation.
To find the inherited action of $U_\rho$, choose an $A$ site as the rotation center and sublattice offsets $\mathbf t_A=0$, $\mathbf t_B=\boldsymbol\delta_1$.
The $120^\circ$ counterclockwise rotation sends $\mathbf K$ to $\mathbf K+\mathbf G$, where $\mathbf G=(-2\pi/\sqrt3,2\pi/3)$ is a reciprocal vector.
In the actual-position Fourier convention, $c_{s,\mathbf k+\mathbf G}=e^{-\mathrm i\mathbf G\cdot\mathbf t_s}c_{s,\mathbf k}$.
The two sublattices thus acquire phases $1$ and $e^{-\mathrm i\mathbf G\cdot\boldsymbol\delta_1}=\omega$, giving $D_\rho=\operatorname{diag}(1,\omega)$.

In this basis, the kinetic matrix is real and the symmetry action factorizes:
\begin{equation}
\begin{aligned}
 U h_M(\mathbf K+\mathbf q)U^\dagger
 &=v(q_x\sigma_3-q_y\sigma_1)+O(q^2),\\
 U D_\rho U^\dagger
 &=e^{-\mathrm i\pi/3}\,\underbrace{e^{-\mathrm i\pi\sigma_2/3}}_{S_\rho}.
\end{aligned}
\label{eq:sm-kekule-basis}
\end{equation}
The two Majoranas have identical kinetic terms.
Since $U\mathbf c_{\mathbf K+\mathbf q}=(\xi_1+\mathrm i\xi_2)/\sqrt2$ and $S_\rho$ is real, the scalar phase $e^{-\mathrm i\pi/3}$ mixes the two flavors by the real rotation $O_\rho$ below, while $S_\rho$ acts on each spinor.
Promoting $\varphi$ to a fluctuating bond-order field and writing $\Xi=(\xi_1,\xi_2)^{\mathsf T}$, we obtain
\begin{equation}
 U_\rho:\quad \Xi\mapsto (O_\rho\otimes S_\rho)\Xi,\qquad
 \varphi\mapsto\omega\varphi,\qquad
 O_\rho=\begin{pmatrix}\frac12&\frac{\sqrt3}{2}\\-\frac{\sqrt3}{2}&\frac12\end{pmatrix}.
\label{eq:sm-kekule-s3}
\end{equation}
The vertical mirror $R_{\mathrm v}:(x,y)\mapsto(-x,y)$ exchanges $\mathbf K\leftrightarrow\mathbf K'$ while preserving the sublattices.
Since $UU^{\mathsf T}=-\mathrm i\sigma_1$, Eq.~\eqref{eq:sm-kekule-fields} gives $\xi_1\mapsto P\xi_2$ and $\xi_2\mapsto P\xi_1$, with $P=-\sigma_1$.
Thus $U_\tau$ sends $\Xi\mapsto (\tau_x\otimes P)\Xi$ and $\varphi\mapsto\varphi^*$, where $\tau_x$ exchanges the flavors.
Coordinate transformations are implicit; $S_\rho^3=O_\rho^3=-1$ ensures $U_\rho^3=1$.

With $\bar\xi_a=\xi_a^{\mathsf T}\sigma_2$, the identity $P^{\mathsf T}\sigma_2P=-\sigma_2$ gives $\bar\xi_1\xi_1\mapsto-\bar\xi_2\xi_2$ and vice versa.
The common mass $\bar\xi_1\xi_1+\bar\xi_2\xi_2$ is therefore forbidden by mirror symmetry.
The bilinears $\mathcal B_1=(\bar\xi_1\xi_1-\bar\xi_2\xi_2)/2$ and $\mathcal B_2=\bar\xi_1\xi_2$ are respectively mirror even and odd, transforming together as $(\operatorname{Re}\varphi,\operatorname{Im}\varphi)$ under $S_3$.
Their invariant contraction $y[\operatorname{Re}\varphi\,\mathcal B_1+\operatorname{Im}\varphi\,\mathcal B_2]$ gives the leading Yukawa coupling~\cite{JianYao2017FIQCP}.
After canonical normalization and setting the fermion velocity to one, the effective Euclidean Lagrangian is
\begin{equation}
 \mathcal L_{\mathrm K}
 =\frac12\bar\xi_a\bigl[\delta_{ab}\gamma^\mu\partial_\mu+\mathsf m_{ab}(\varphi)\bigr]\xi_b
 +|\partial_\tau\varphi|^2+v_\varphi^2|\nabla\varphi|^2+V_{\mathrm K}(\varphi),
\label{eq:sm-kekule-yukawa}
\end{equation}
with $a,b=1,2$ summed and $V_{\mathrm K}=r_\varphi|\varphi|^2+u_\varphi|\varphi|^4+w_\varphi(\varphi^3+\varphi^{*3})$.
For real $y$, the Yukawa mass matrix and its eigenvalues are
\begin{equation}
 \mathsf m(\varphi)=y
 \begin{pmatrix}
 \operatorname{Re}\varphi&\operatorname{Im}\varphi\\
 \operatorname{Im}\varphi&-\operatorname{Re}\varphi
 \end{pmatrix},
 \qquad m_\pm=\pm y|\varphi|.
\label{eq:sm-kekule-masses}
\end{equation}

For $\langle\varphi\rangle\ne0$, the time-reversal-preserving Kekul{\'e} mass gaps both Majoranas; the resulting band has zero Chern number and $c_-=0$.
The ungauged fermion band can be continuously deformed to isolated single-color dimers without closing its gap, identifying a topologically trivial superconductor~\cite{Quinn2015Kekule}.
The cubic anisotropy locks three symmetry-related values of $\langle\varphi\rangle$, breaking $S_3$ to $\mathbb Z_2$ while preserving $\mathrm T$.
Since $\varphi$ is gauge neutral, its expectation value leaves the $\mathbb Z_2$ gauge structure intact.
Restoring the deconfined gauge field, equivalently gauging fermion parity, therefore gives TC order in each vacuum~\cite{Kitaev2006}.
This yields $\bigoplus_{i=1}^3\mathrm{TC}_i$, the same proposed intermediate phase approached from the color-code side.

\section{Representation theory and Chern--Simons theory of \texorpdfstring{$\mathrm{PSU}(3)$ and $\mathrm{Spin}(8)$}{PSU(3) and Spin(8)}}
\label{sec:sm-spin8-psu3}

We collect the representation-theoretic facts underlying the CSH construction, using the Hermitian gauge fields and trace normalizations of the main text.
They fix both the charge-two Abelian restriction of the $\mathrm{Spin}(8)$ vector and the factor of three in the Chern--Simons level under the adjoint embedding.

The Lie algebras $\mathfrak{spin}(8)$ and $\mathfrak{su}(3)$ have the Dynkin diagrams~\cite{Slansky1981}
\begin{equation}
\begin{array}{c@{\hspace{2.2cm}}c}
\begin{tikzpicture}[
 baseline=(current bounding box.center),
 root/.style={circle,draw=black,fill=white,inner sep=0pt,
              minimum size=5.5pt,line width=0.5pt},
 bond/.style={line width=0.5pt}]
 \node[root,label={[font=\small]left:$\alpha_1$}] (a1) at (180:0.95) {};
 \node[root,label={[font=\small]above left:$\alpha_2$}] (a2) at (0,0) {};
 \node[root,label={[font=\small]above right:$\alpha_3$}] (a3) at (60:0.95) {};
 \node[root,label={[font=\small]below right:$\alpha_4$}] (a4) at (-60:0.95) {};
 \draw[bond] (a2)--(a1) (a2)--(a3) (a2)--(a4);
\end{tikzpicture}
&
\begin{tikzpicture}[
 baseline=(current bounding box.center),
 root/.style={circle,draw=black,fill=white,inner sep=0pt,
              minimum size=5.5pt,line width=0.5pt},
 bond/.style={line width=0.5pt}]
 \node[root,label={[font=\small]below:$\beta_1$}] (b1) at (-0.475,0) {};
 \node[root,label={[font=\small]below:$\beta_2$}] (b2) at (0.475,0) {};
 \draw[bond] (b1)--(b2);
\end{tikzpicture}
\\[5pt]
D_4:\ \mathfrak{spin}(8)&A_2:\ \mathfrak{su}(3)
\end{array}
\label{eq:sm-dynkin-diagrams}
\end{equation}
The outer $D_4$ nodes $\alpha_1,\alpha_3,\alpha_4$ correspond to the vector $\mathbf8_v$ and the two spinors $\mathbf8_s,\mathbf8_c$, respectively.

\paragraph*{Center and Cartan restriction.}
The vector $\mathbf8_v$ and the two chiral spinors $\mathbf8_s,\mathbf8_c$ are inequivalent real representations, distinguished by their central characters~\cite{Slansky1981}.
Write $Z(\mathrm{Spin}(8))=\{1,z_v,z_s,z_c\}\cong\mathbb Z_2\times\mathbb Z_2$, with $z_vz_s=z_c$ and $\ker R_x=\{1,z_x\}$ for $x=v,s,c$.
Their center actions are
\begin{equation}
\begin{array}{c|ccc}
 &z_v&z_s&z_c\\ \hline
\mathbf8_v&+1&-1&-1\\
\mathbf8_s&-1&+1&-1\\
\mathbf8_c&-1&-1&+1
\end{array}.
\label{eq:sm-spin8-center}
\end{equation}
In particular, $z_v=-1\in\mathrm{Spin}(8)$ generates the kernel of the covering map to $\mathrm{SO}(8)$.
Triality permutes the three representations together with these central elements.

Choose the Cartan $U(1)$ generated by rotations in one coordinate plane; it contains the center subgroup $\mathbb Z_2^v=\{1,z_v\}$.
In complex weight bases, the same generator $T_z$ is represented by $R_{s,c}(T_z)=(\sigma_z/2)\otimes\mathbf1_4$ on either spinor, but by $R_v(T_z)=\sigma_z\oplus0_6$ on the vector.
Thus $e^{\mathrm i\theta T_z}$ has period $4\pi$ in $\mathrm{Spin}(8)$, although its vector action has period $2\pi$; writing $\theta=2\alpha$ with $\alpha\sim\alpha+2\pi$ gives the integral charge generator $Q=2T_z$ and hence vector charges $\pm2$.
The branching rules for the complexified representations under $U(1)_Q$ are
\begin{equation}
\begin{aligned}
 \mathbf8_v
 &\longrightarrow\mathbf1_{+2}\oplus\mathbf1_{-2}\oplus6\cdot\mathbf1_0,\\
 \mathbf8_{s,c}
 &\longrightarrow4\cdot\mathbf1_{+1}\oplus4\cdot\mathbf1_{-1}.
\end{aligned}
\label{eq:sm-cartan-branching}
\end{equation}
Here $\mathbf1_q$ denotes a one-dimensional representation of charge $q$.
The halfway element $e^{\mathrm i\pi Q}=z_v$ acts trivially on the vector and as $-1$ on both spinors, as required.

For a compact connection $a_v$ with holonomy defined modulo $2\pi$, restrict $b=Q a_v$.
Since $\operatorname{Tr}_v Q^2=8$, the Chern--Simons term becomes
\begin{equation}
 \left.k\,\operatorname{CS}_{\mathfrak{spin}(8)}[b]\right|_{b=Q a_v}
 =\frac{k}{8\pi}\operatorname{Tr}_v(Q^2)\,a_v\wedge da_v
 =\frac{4k}{4\pi}a_v\wedge da_v.
\label{eq:sm-cartan-cs}
\end{equation}
The spectator at $k=-1$ therefore induces $U(1)_{-4}$, with a $z_v$ holonomy represented by $\pi$ flux.
The Chern--Simons Gauss law attaches electric charge $q=2$ to this flux, giving topological spin $\theta=\exp[\mathrm i\pi q^2/(-4)]=-1$.
This restriction evaluates the inherited topological action and does not require an intermediate $U(1)$ phase.

\paragraph*{Adjoint embedding and charge conjugation.}
Define $B(X,Y)=\operatorname{Tr}_{\mathrm{adj}}(\operatorname{ad}_X\operatorname{ad}_Y)$, where $\operatorname{ad}_X(Y)=[X,Y]$ acts on the complexified algebra.
The Jacobi identity gives $[\operatorname{ad}_Z,\operatorname{ad}_X]=\operatorname{ad}_{[Z,X]}$; trace cyclicity then implies $B([Z,X],Y)+B(X,[Z,Y])=0$.
Thus the adjoint action preserves the positive-definite inner product $B$ on the eight-dimensional real space of traceless Hermitian matrices, so it acts by orthogonal matrices in a $B$-orthonormal basis.
Since $\mathrm{SU}(3)$ is connected, these matrices have determinant $+1$ and hence lie in $\mathrm{SO}(8)$; the kernel $\mathbb Z_3$ then gives a faithful map $\mathrm{PSU}(3)\to\mathrm{SO}(8)$.
The induced map $\pi_1(\mathrm{PSU}(3))=\mathbb Z_3\to\pi_1(\mathrm{SO}(8))=\mathbb Z_2$ is necessarily trivial, so this map lifts to an injective homomorphism $\iota:\mathrm{PSU}(3)\hookrightarrow\mathrm{Spin}(8)$~\cite{KomargodskiAdjoint2021}.
Its differential $\iota_*$ acts on Lie-algebra generators and gauge fields, with $\iota(e^{\mathrm iX})=e^{\mathrm i\iota_*(X)}$; below we write its image as matrices in the vector representation.
Under this embedding, the finite-dimensional representations branch as~\cite{Slansky1981}
\begin{equation}
 \mathbf8_v,\ \mathbf8_s,\ \mathbf8_c\ \longrightarrow\ \mathbf8_{\mathrm{adj}},
 \qquad
 \mathbf{28}\ \longrightarrow\ \mathbf8_{\mathrm{adj}}\oplus\mathbf{10}\oplus\overline{\mathbf{10}}.
\label{eq:sm-adjoint-branching}
\end{equation}
This special adjoint embedding must be distinguished from the regular embedding, for which $\mathbf8_v\to\mathbf3\oplus\overline{\mathbf3}\oplus2\cdot\mathbf1$.

In the Hermitian convention, $\mathrm{PSU}(3)$ charge conjugation acts as $X\mapsto-X^{\mathsf T}$~\cite{Cordova2018}.
On the real adjoint space its matrix $C\in\mathrm O(8)$ has eigenvalue $-1$ on the five real symmetric traceless generators and $+1$ on the three imaginary antisymmetric generators, giving $C^2=\mathbf1_8$ and $\det C=-1$.
Write $\mathcal C_3([g])=[g^*]$ for this automorphism of $\mathrm{PSU}(3)$, and $\mathcal C_8$ for the automorphism of $\mathrm{Spin}(8)$ induced by $C$.
The identity $C\iota_*(X)C^{\mathsf T}=\iota_*(-X^{\mathsf T})$ integrates to the commuting diagram
\begin{equation}
\begin{tikzpicture}[baseline=(current bounding box.center),
 every node/.style={inner sep=3pt},line width=0.45pt]
 \node (p1) at (0,0) {$\mathrm{PSU}(3)$};
 \node (s1) at (3,0) {$\mathrm{Spin}(8)$};
 \node (p2) at (0,-1.15) {$\mathrm{PSU}(3)$};
 \node (s2) at (3,-1.15) {$\mathrm{Spin}(8)$};
 \draw[->] (p1)--node[above] {$\iota$} (s1);
 \draw[->] (p2)--node[above] {$\iota$} (s2);
 \draw[->] (p1)--node[left] {$\mathcal C_3$} (p2);
 \draw[->] (s1)--node[right] {$\mathcal C_8$} (s2);
\end{tikzpicture}
\qquad
\mathcal C_8\circ\iota=\iota\circ\mathcal C_3.
\label{eq:sm-conjugation-diagram}
\end{equation}
The action on the center identifies which triality reflection $\mathcal C_8$ implements.
Choose Euclidean Clifford generators satisfying $\{\Gamma_I,\Gamma_J\}=2\delta_{IJ}$ and define the volume element $\Omega=\Gamma_1\cdots\Gamma_8$, with $\Omega^2=1$.
Taking $\mathbf8_s$ to have positive chirality gives $z_s=\Omega$ and $z_c=-\Omega$, while $z_v=-1$ is the scalar element generating the kernel of $\mathrm{Spin}(8)\to\mathrm{SO}(8)$.
The Clifford-algebra automorphism induced by $C$ fixes this scalar but sends $\Omega\mapsto(\det C)\Omega=-\Omega$, hence
\begin{equation}
 \mathcal C_8(z_v)=z_v,\qquad
 \mathcal C_8(z_s)=z_c,\qquad
 \mathcal C_8(z_c)=z_s.
\label{eq:sm-conjugation-center}
\end{equation}
By the central characters in Eq.~\eqref{eq:sm-spin8-center}, $R_s\circ\mathcal C_8$ has the character of $R_c$, and conversely, while that of $R_v$ is unchanged.
Thus $\mathcal C_8$ fixes $\mathbf8_v$ and exchanges $\mathbf8_s\leftrightarrow\mathbf8_c$~\cite{CordovaOrthogonal2018}.
Together with Eq.~\eqref{eq:sm-conjugation-diagram}, this gives the combined symmetry $\mathcal C$ of the main text; the spinor exchange is a triality reflection, not complex conjugation of the real spinor representations.
Writing $a=a^A\lambda_A/2$ in terms of the Gell--Mann matrices $\lambda_A$, the combined symmetry acts in compatible bases as $(a_{\mathrm{adj}},b_v,\Phi)\mapsto(Ca_{\mathrm{adj}}C^{\mathsf T},Cb_vC^{\mathsf T},C\Phi C^{\mathsf T})$.

\paragraph*{Chern--Simons levels.}
Take $[t^a,t^b]=\mathrm i f^{ab}{}_c t^c$ and $\operatorname{Tr}_{\mathbf3}(t^at^b)=\delta^{ab}/2$, so that $f^{aCD}f^{bCD}=3\delta^{ab}$.
The embedded generators are $[\iota_*(t^a)]_{BC}=-\mathrm i f^{aBC}$, and hence
\begin{equation}
 \frac12\operatorname{Tr}_v\!\left[\iota_*(t^a)\iota_*(t^b)\right]
 =\frac32\delta^{ab}
 =3\operatorname{Tr}_{\mathbf3}(t^at^b).
\label{eq:sm-embedding-index}
\end{equation}
The factor $1/2$ is the one in the $\mathrm{Spin}(8)$ Chern--Simons normalization relative to the vector trace; the embedding index is therefore three, rather than the raw trace ratio six.
Since $\iota_*$ preserves the Lie bracket, the same factor multiplies the cubic term, giving
\begin{equation}
 \operatorname{CS}_{\mathfrak{spin}(8)}[\iota_*(a)]
 =3\operatorname{CS}_{\mathfrak{su}(3)}[a],
 \qquad k_{\mathrm{PSU}(3)}=3k_{\mathrm{Spin}(8)}.
\label{eq:sm-level-restriction}
\end{equation}
Here the $\mathrm{PSU}(3)$ level uses the local fundamental-trace convention of the main text, with its global definition inherited through $\iota$.
In particular, the spectator level $-1$ becomes $-3$ on the locked connection and cancels the original level $+3$.

\paragraph*{Free-fermion check.}
The factor of three also follows directly from eight chiral Majorana fermions $\psi^A$, whose bilinears generate $\widehat{\mathfrak{so}}(8)_1$~\cite{KomargodskiAdjoint2021}.
Under the adjoint embedding these fermions form an $\mathrm{SU}(3)$ adjoint, so the embedded currents are
\begin{equation}
 \psi^A(z)\psi^B(w)\sim\frac{\delta^{AB}}{z-w},
 \qquad
 j^a=-\frac{\mathrm i}{2}f^{aBC}:\!\psi^B\psi^C\!:,
\label{eq:sm-fermion-currents}
\end{equation}
where colons denote normal ordering.
Wick contraction gives
\begin{equation}
 j^a(z)j^b(w)\sim
 \frac{3\delta^{ab}/2}{(z-w)^2}
 +\frac{\mathrm i f^{ab}{}_c j^c(w)}{z-w}.
\label{eq:sm-current-ope}
\end{equation}
The double pole is $\tfrac12 f^{aCD}f^{bCD}=3\delta^{ab}/2$; comparing with the standard coefficient $k\delta^{ab}/2$ gives level $k=3$.
Thus the free-fermion currents realize $\widehat{\mathfrak{su}}(3)_3\subset\widehat{\mathfrak{so}}(8)_1$, providing a direct current-algebra check of Eq.~\eqref{eq:sm-level-restriction}.

\section{Higgs phase and residual topological order}
\label{sec:sm-higgs}

\paragraph*{Unbroken gauge group.}
For $\langle\Phi\rangle=v_{\mathrm H}\mathbf1_8$, a gauge transformation $(g,h)\in\mathrm{PSU}(3)\times\mathrm{Spin}(8)$ preserves the condensate precisely when $R_{\mathrm{adj}}(g)R_v(h)^{\mathsf T}=\mathbf1_8$.
The embedding of Sec.~\ref{sec:sm-spin8-psu3} obeys $R_v(\iota(g))=R_{\mathrm{adj}}(g)$, so $h=\iota(g)z_v^\epsilon$ with $\epsilon=0,1$.
The stabilizer is therefore
\begin{equation}
 H=\bigl\{(g,\iota(g)z_v^\epsilon)\mid g\in\mathrm{PSU}(3),\ \epsilon=0,1\bigr\}
 \cong\mathrm{PSU}(3)_{\mathrm d}\times\mathbb Z_2^v.
\label{eq:sm-higgs-stabilizer}
\end{equation}
Here $\mathrm{PSU}(3)_{\mathrm d}=\{(g,\iota(g))\}$ is the diagonal subgroup.
The two factors commute and have trivial intersection, so no further quotient is required.
The discrete factor is invisible to the vector Higgs but acts on spinor Wilson lines.
On the connected component, $a=a_{\mathrm d}$ and $b=\iota_*(a_{\mathrm d})$; Eq.~\eqref{eq:sm-level-restriction} then gives net level $3-3=0$, leaving $\mathrm{PSU}(3)$ Yang--Mills theory.

\paragraph*{Inherited discrete action.}
To determine the action of the remaining $\mathbb Z_2^v$ gauge field, use the Cartan restriction of Eq.~\eqref{eq:sm-cartan-cs}: the level-$-1$ spectator gives $U(1)_{-4}$, and a charged vector component has charge two.
Write this component as $\phi_2=v_2e^{\mathrm i\vartheta}$ with $\vartheta\sim\vartheta+2\pi$.
Its phase stiffness is $\tfrac{v_2^2}{2}(d\vartheta-2a_v)\wedge*(d\vartheta-2a_v)$; gauge transformations $\vartheta\mapsto\vartheta+2\alpha$ leave the condensate invariant for $\alpha=0,\pi$, retaining precisely $\{1,z_v\}$.

Dualizing the compact phase introduces a two-form current $j$ coupled through $-\mathrm i j\wedge(d\vartheta-2a_v)$~\cite{KapustinSeiberg2014}.
Integrating over the smooth part of $\vartheta$ imposes $dj=0$; choose $j=-d\beta/(2\pi)$, where $\beta$ is a compact $U(1)$ connection.
The coupling $2\mathrm i j\wedge a_v$ then becomes $-2\mathrm i\beta\wedge da_v/(2\pi)$ up to a total derivative.
The quadratic current term gives a Maxwell term for $\beta$, which is subleading in the infrared.
Combining the BF term with the inherited Chern--Simons term yields
\begin{equation}
 \mathcal L_K=\frac{4\mathrm i}{4\pi}a_v\wedge da_v
 -\frac{2\mathrm i}{2\pi}\beta\wedge da_v
 =-\frac{\mathrm i}{4\pi}K_{IJ}a_I\wedge da_J,
 \qquad
 K=\begin{pmatrix}-4&2\\2&0\end{pmatrix},
\label{eq:sm-higgs-k}
\end{equation}
with $(a_1,a_2)=(a_v,\beta)$, as in the main text.
Compactness is essential: the BF constraint retains $a_v$ holonomies $0$ or $\pi$ modulo $2\pi$, rather than eliminating the gauge field.

The integral change of variables
\begin{equation}
 \begin{pmatrix}a_v\\\beta\end{pmatrix}
 =W\begin{pmatrix}\widetilde a\\\widetilde\beta\end{pmatrix},
 \qquad
 W=\begin{pmatrix}1&0\\1&1\end{pmatrix},
 \qquad
 W^{\mathsf T}KW=\begin{pmatrix}0&2\\2&0\end{pmatrix}
 \equiv K_{\mathrm{TC}}
\label{eq:sm-higgs-tc}
\end{equation}
preserves the compact charge lattice because $\det W=1$~\cite{WenZee1992}.
Thus the residual theory is the untwisted toric code: its two bosonic generators braid with phase $-1$, and its four topological spins are $\{1,1,1,-1\}$.
The double-semion theory instead has spins $\{1,1,\mathrm i,-\mathrm i\}$ and is not equivalent to this $K$ matrix.
In the original basis, the $\beta$ Wilson line has charge vector $\ell=(0,1)$, carries $\pi$ flux of $a_v$, and has spin $\theta_\ell=e^{\mathrm i\pi\ell^{\mathsf T}K^{-1}\ell}=-1$.
It is the toric-code fermion carrying the inherited one-form symmetry.

\section{Gauging triality and the inverse anyon condensation}
\label{sec:sm-triality-gauging}

We explain the two changes of description used in the main text: gauging cyclic triality combines $v,s,c$ into one non-Abelian anyon, while condensing the resulting gauge-charge lines restores the three channels~\cite{Teo2015,Barkeshli2019}.
The $\mathrm{Spin}(8)_{-1}$ factor remains a spectator throughout.

\paragraph*{Gauging the zero-form $\mathbb Z_3$.}
The symmetry $\rho:v\mapsto s\mapsto c\mapsto v$ permutes the three nontrivial anyons of $\mathrm{Spin}(8)_1$, each of dimension one and spin $-1$.
Gauging also requires a choice of discrete torsion in $H^3(\mathbb Z_3,U(1))=\mathbb Z_3$, equivalently a choice of $\mathbb Z_3$ SPT stacked before gauging.
This changes the statistics of the dynamical $\mathbb Z_3$ flux lines but leaves the local critical data, including the scaling dimensions and correlation functions of gauge-invariant local operators, unchanged~\cite{Ji2026,LuWangVishwanath}.
We choose the class giving $\mathrm{SU}(3)_3$, corresponding to $m=2$ in the conventions of Ref.~\cite{Teo2015}.

To see the resulting anyon content, denote the $\rho$ and $\rho^{-1}$ defects by $\sigma^+$ and $\sigma^-$.
Write the Abelian fusion group as $\mathsf A=\{0,v,s,c\}\simeq\mathbb Z_2\times\mathbb Z_2$ in additive notation, with $0$ denoting the vacuum.
Creating a $b,-b$ pair near $\sigma^+$ and winding $b$ once around it changes $b$ into $\rho(b)$; recombining the pair at the defect attaches $\rho(b)-b$ without changing its species.
Attachments are therefore classified by $\mathsf A/(1-\rho)\mathsf A$~\cite{Teo2015}.
Here $(1-\rho)v=c$, $(1-\rho)s=v$, and $(1-\rho)c=s$, so $(1-\rho)\mathsf A=\mathsf A$ and the quotient is trivial.
The same holds for $\rho^{-1}$: each orientation has a single defect species, with $a\times\sigma^\pm\simeq\sigma^\pm$ for every $a\in\{\mathbf1,v,s,c\}$.
Each absorbed anyon gives one fusion channel of the defect with its antidefect, hence
\begin{equation}
 \sigma^+\times\sigma^-=\mathbf1\oplus v\oplus s\oplus c,
 \qquad d_{\sigma^\pm}^2=d_{\mathbf1}+d_v+d_s+d_c=4.
\label{eq:sm-triality-defect-fusion}
\end{equation}
Thus $d_{\sigma^\pm}=2$, reflecting the growth of the fusion space of many defects.
Fusing two defects of the same orientation composes their twists as $\rho^2=\rho^{-1}$, giving $\sigma^+\times\sigma^+=2\sigma^-$ and similarly with $+\leftrightarrow-$.
The multiplicity two counts independent fusion states of the same outgoing defect, distinguishable by Wilson-loop measurements~\cite{Teo2015}.
After gauging, each flux can carry one of three $\mathbb Z_3$ charges, yielding six dimension-two anyons $\sigma_n^\pm$ with $n=0,1,2$.
The invariant vacuum gives the three electric charges $\mathbf1,J,J^2$, whereas the orbit $\{v,s,c\}$ becomes a single dimension-three anyon $\Psi$.
For the chosen SPT class, the flux weights are $h_{\sigma_n^\pm}=(2+3n)/9\pmod1$.
The complete spectrum matches $\mathrm{SU}(3)_3$ as follows; $(p,q)$ are Dynkin labels and $\theta=e^{2\pi\mathrm i h}$.
\begin{center}
\renewcommand{\arraystretch}{1.15}
\setlength{\tabcolsep}{10pt}
\begin{tabular}{lccc}
\toprule
Gauged sector & $\mathrm{SU}(3)_3$ labels $(p,q)$ & $d$ & $h\pmod1$\\
\midrule
Vacuum $\mathbf1$ & $(0,0)$ & $1$ & $0$\\
Electric charges $J,J^2$ & $(3,0),(0,3)$ & $1$ & $0$\\
Orbit $\Psi=\mathbf8_{\mathrm{adj}}$ & $(1,1)$ & $3$ & $1/2$\\
Fluxes $\sigma_0^+,\sigma_0^-$ & $(0,1),(1,0)$ & $2$ & $2/9$\\
Fluxes $\sigma_1^+,\sigma_1^-$ & $(2,0),(0,2)$ & $2$ & $5/9$\\
Fluxes $\sigma_2^+,\sigma_2^-$ & $(1,2),(2,1)$ & $2$ & $8/9$\\
\bottomrule
\end{tabular}
\end{center}
In particular, the total quantum dimension satisfies $\mathcal D_{\mathrm{SU}(3)_3}^2=3\cdot1^2+3^2+6\cdot2^2=36$, so gauging multiplies $\mathcal D_{\mathrm{Spin}(8)_1}=2$ by $|\mathbb Z_3|=3$.

\paragraph*{Condensing the dual one-form $\mathbb Z_3^{(1)}$.}
The electric charge $J=(3,0)$ obeys $J^3=\mathbf1$ and acts on integrable weights, $p,q\geq0$ with $p+q\leq3$, by
\begin{equation}
 J\times(p,q)=(3-p-q,p),
 \qquad
 h_{(p,q)}=\frac{p^2+q^2+pq+3p+3q}{18}.
\label{eq:sm-su3-simple-current}
\end{equation}
Both $J$ and $J^2$ have $h=1$ and trivial mutual braiding, so they generate a nonanomalous one-form symmetry.
Gauging it means condensing the algebra $\mathcal A=\mathbf1\oplus J\oplus J^2$~\cite{HsinLamSeiberg2019}.
An anyon remains deconfined only if its full braiding around the condensate is trivial.
Using Eq.~\eqref{eq:sm-su3-simple-current}, this phase is
\begin{equation}
 M_{J,(p,q)}
 =\frac{\theta_{J\times(p,q)}}{\theta_J\theta_{(p,q)}}
 =e^{-2\pi\mathrm i(p+2q)/3}
 =\omega^{p+2q},
 \qquad \omega=e^{-2\pi\mathrm i/3}.
\label{eq:sm-su3-monodromy}
\end{equation}
Thus all six flux anyons are confined, while $\{\mathbf1,J,J^2\}$ is identified with the vacuum.
The only other unconfined sector is $\Psi=(1,1)$, a fixed point of fusion with $J$.
It resolves into the three characters of its $\mathbb Z_3$ stabilizer, giving
\begin{equation}
 \Psi\longrightarrow v\oplus s\oplus c,
 \qquad d_v=d_s=d_c=1,
 \qquad \theta_v=\theta_s=\theta_c=-1.
\label{eq:sm-adjoint-resolution}
\end{equation}
The surviving lines have $\mathbb Z_2\times\mathbb Z_2$ fusion, with $v^2=s^2=c^2=\mathbf1$ and $v\times s=c$ cyclically: a fermionic generator of $\mathbb Z_4$ would have a bosonic square, incompatible with these spins.
Their total squared dimension is $36/(\dim\mathcal A)^2=4$, identifying the daughter theory as $\mathrm{Spin}(8)_1$.

The dual magnetic $\mathbb Z_3$ acts on $\mathcal A$ by $J^r\mapsto\omega^rJ^r$.
This shifts the stabilizer character of a resolved line and hence cyclically permutes $v,s,c$, recovering triality.
The two operations are therefore inverse~\cite{Tachikawa2020}:
\begin{equation}
 \mathrm{Spin}(8)_1
 \xrightarrow{\text{gauge }\mathbb Z_3}\mathrm{SU}(3)_3,
 \qquad
 \mathrm{SU}(3)_3/\mathbb Z_3^{(1)}
 =\mathrm{PSU}(3)_3\simeq\mathrm{Spin}(8)_1.
\label{eq:sm-inverse-gauging}
\end{equation}
Here the first arrow uses the SPT choice specified above; when backgrounds for the restored $\mathbb Z_3$ are retained, the inverse operation recovers $\mathrm{Spin}(8)_1$ together with the chosen SPT response.
In the main-text construction, this inverse operation recovers the original three condensation channels from the single adjoint channel.

\section{Majorana regularization and the boson-fermion duality}
\label{sec:sm-qcd-regulator}

We use Euclidean signature with $Z=\int\mathcal D(\cdots)e^{-S_E}$ and $S_E=\int d^3x\,\mathcal L$.
In an oriented orthonormal frame, take $\gamma^\mu=\sigma_\mu$ ($\mu=1,2,3$), $\epsilon^{123}=+1$, and $\bar\chi^A=(\chi^A)^{\mathsf T}\sigma_2$.

In the QCD and Yang--Mills terms we suppress the trace labels $\mathbf3$ for $\chi,a$ and $v$ for $b$.
The normalization $\operatorname{Tr}_{\mathbf3}(t^At^B)=\delta^{AB}/2$ supplies the Majorana factor $1/2$.

Gauge connections are Hermitian, with $f_a=da-\mathrm i a\wedge a$ and $f_a^2=f_{a,\mu\nu}f_a^{\mu\nu}$, and similarly for $b$.
In real adjoint and vector bases, $-\mathrm i a_{\mathrm{adj}}$ and $-\mathrm i b_v$ are real antisymmetric matrices, so $D_\mu\Phi$ is real.
The Chern--Simons forms are
$
 \operatorname{CS}_{\mathfrak{su}(3)}[a]
 =\frac{1}{4\pi}\operatorname{Tr}_{\mathbf{3}}
\left(a\,da-\frac{2\mathrm i}{3}a^3\right),
 \operatorname{CS}_{\mathfrak{spin}(8)}[b]
 =\frac{1}{8\pi}\operatorname{Tr}_{v}
\left(b\,db-\frac{2\mathrm i}{3}b^3\right)$.
We identify these three-forms with their scalar densities in component Lagrangians.
Level $k$ contributes $-\mathrm i k\int\operatorname{CS}$ to $S_E$, so the levels $(+3,-1)$ in Eq.~\eqref{eq:csh} correspond to $-3\mathrm i\operatorname{CS}[a]+\mathrm i\operatorname{CS}[b]$.
We denote the real response phase by $I_{\mathrm{top}}$, with $Z_{\mathrm{top}}=e^{\mathrm iI_{\mathrm{top}}}$ and $S_{E,\mathrm{top}}=-\mathrm iI_{\mathrm{top}}$.

\paragraph*{Regularization scheme.}
With the mass term $+m\operatorname{Tr}(\bar\chi\chi)$, we choose a positive Pauli--Villars (PV) reference mass $M_{\mathrm{PV}}>0$.
On an oriented Riemannian spin three-manifold $Y$, define the self-adjoint $D_R=-\mathrm i\gamma^\mu D_\mu$, with $D_\mu=\nabla_\mu^{\mathrm{spin}}-\mathrm i A_{R,\mu}$ and background connection $A_R$ in representation $R$.
The fermion kernel is $\gamma^\mu D_\mu+m=m+\mathrm iD_R$, giving, after parity-even subtractions~\cite{Witten2016Fermion},
\begin{equation}
 Z_R(m)=\lim_{M_{\mathrm{PV}}\to+\infty}
 \left[\prod_\lambda\frac{m+\mathrm i\lambda}{M_{\mathrm{PV}}+\mathrm i\lambda}\right]_{\mathrm{PV}}^{1/2},
 \qquad
 Z_R(0)=|\operatorname{Pf}|\,e^{\mathrm i\pi\eta_R/4}.
\label{eq:sm-pv-majorana}
\end{equation}
The PV prescription fixes the square-root branch continuously.
Here $\eta_R$ is the analytic continuation to $s=0$ of $\sum_{\lambda\ne0}\operatorname{sgn}(\lambda)|\lambda|^{-s}$ over the full complex Dirac spectrum, including degeneracies; the factor $1/4$ already accounts for the Majorana condition.
We assume no zero modes; when they occur, the massless Pfaffian vanishes.

For a neutral cone, $m>0$ is the trivial reference phase, while $m<0$ has response $e^{\mathrm i\pi\eta_{\mathbf1}/2}=e^{\mathrm iI_{\mathrm g}}$, with $I_{\mathrm g}=\int_Y\operatorname{CS}_{\mathrm g}[\varpi]$.
Here $\operatorname{CS}_{\mathrm g}[\varpi]=\frac1{192\pi}\operatorname{tr}\left[\varpi\,d\varpi+\frac23\varpi^3\right]$, where $\varpi$ is the spin connection~\cite{Seiberg2016DualityWeb}.
In our framing convention $Z_{\mathrm{SO}(n)_1}=e^{-\mathrm inI_{\mathrm g}}$, a phase $e^{\mathrm i\ell_{\mathrm g}I_{\mathrm g}}$ carries $c_-^{\mathrm{inv}}=-\ell_{\mathrm g}/2$~\cite{AharonyOrthogonal2017,HsinSeiberg2016}.
Thus $m<0$ is $p-\mathrm i p$ with $c_-=-1/2$, whose $\mathcal{CT}$ conjugate is $p+\mathrm i p$.
These signs fix the PV reference; changing only the latter shifts both massive responses by the same quantized term.

\paragraph*{Adjoint level jump.}
For a $\mathrm{PSU}(3)$ bundle $P$ with background connection $A$, we apply the same prescription to the Dirac operator coupled to $E_{\mathrm{adj}}=P\times_{\mathrm{Ad}}\mathfrak{su}(3)$.
This defines $\eta_{\mathrm{adj}}[A]$ globally even when $P$ does not lift to $\mathrm{SU}(3)$.
Using $\operatorname{Tr}_{\mathrm{adj}}(t^at^b)=3\delta^{ab}=6\operatorname{Tr}_{\mathbf3}(t^at^b)$ and $\dim(\mathrm{adj})=8$, the Atiyah--Patodi--Singer (APS) relation gives~\cite{AharonyOrthogonal2017}
\begin{equation}
 \delta\!\left(\frac\pi4\eta_{\mathrm{adj}}\right)
 =\delta\!\left(\frac32 I_{\mathrm{CS}}[A]+4I_{\mathrm g}\right),
 \qquad I_{\mathrm{CS}}[A]=\int_Y\operatorname{CS}_{\mathfrak{su}(3)}[A].
\label{eq:sm-adjoint-contact}
\end{equation}
The two massive responses are therefore
\begin{equation}
 I_{\mathrm{top}}(m)
 =\frac{1-\operatorname{sgn}m}{2}\bigl(3I_{\mathrm{CS}}[A]+8I_{\mathrm g}\bigr).
\label{eq:sm-adjoint-level-jump}
\end{equation}
The levels $0$ and $3$ have midpoint $3/2$, giving the notation $\mathrm{PSU}(3)_{3/2}$~\cite{Gomis2018}; this contact term is included in the regulated Pfaffian, with no additional half-level Chern--Simons term.
Promoting $A$ to the dynamical connection $a$ gives the branches in Eq.~\eqref{eq:qcd-phase-flows}: $\mathrm{PSU}(3)$ Yang--Mills theory for $m>0$, and $\mathrm{PSU}(3)_3\boxtimes[p-\mathrm i p]^{\otimes8}$ for $m<0$.
In the latter, the eight superconducting layers supply $8I_{\mathrm g}$ and $c_-^{\mathrm{inv}}=-4$, canceling the gauge contribution $c_-=4$.

\paragraph*{Duality with a fixed background.}
Here we spell out the duality to free Majorana fermions used in the main text, including the background counterterm and the matching of the two gapped phases.
With $A$ a fixed $\mathrm{PSU}(3)$ background and $b$ a dynamical $\mathrm{SO}(8)$ connection, the two Euclidean actions are~\cite{Metlitski2017Majorana,AharonyOrthogonal2017,ChenZimet2018}
\begin{equation}
\begin{aligned}
 S_B[\Phi,b;A,r]
 &={}\int_Y d^3x\sqrt g\,\left\{
 \operatorname{Tr}[(D^\mu\Phi)^{\mathsf T}D_\mu\Phi]
 +r\operatorname{Tr}(\Phi^{\mathsf T}\Phi)
 +\mathcal V(\Phi)
 +\frac1{g_b^2}\operatorname{Tr}f_b^2\right\}\\
 &\quad+\mathrm i\int_Y\operatorname{CS}_{\mathfrak{spin}(8)}[b]
 -3\mathrm i\int_Y\operatorname{CS}_{\mathfrak{su}(3)}[A],\\
 S_F[\chi;A,m]
 &={}\int_Y d^3x\sqrt g\,
 \operatorname{Tr}\!\left[\bar\chi\gamma^\mu
 (\nabla_\mu\chi-\mathrm i[A_\mu,\chi])+m\bar\chi\chi\right].
\end{aligned}
\label{eq:sm-background-dual-actions}
\end{equation}
Here $D_\mu\Phi=\partial_\mu\Phi-\mathrm iA_{\mathrm{adj},\mu}\Phi+\mathrm i\Phi b_{v,\mu}$, and $\mathcal V$ includes $\det\Phi$, which removes the extra independent reflection $\mathbb Z_2$ and selects a single Higgs orientation~\cite{ChenZimet2018}.
The fermionic path integral is regulated by dividing by the positive-mass PV Pfaffian as in Eq.~\eqref{eq:sm-pv-majorana}; no separate half-level Chern--Simons term is added.
In this convention the background-field duality reads
\begin{equation}
 \int\mathcal Db\,\mathcal D\Phi\,e^{-S_B[\Phi,b;A,r]}
 \underset{\mathrm{IR}}{\simeq}
 \left[\int\mathcal D\chi\,e^{-S_F[\chi;A,m]}\right]_{\mathrm{PV},\,M_{\mathrm{PV}}>0},
 \qquad m\propto-r.
\label{eq:sm-background-bosonization}
\end{equation}
The parent duality matches the connected $\mathrm{SO}(8)$ flavor rotations, restricted here to the adjoint $\mathrm{PSU}(3)$ subgroup, and maps the bosonic gauge theory's magnetic $\mathbb Z_2$ to fermion parity $(-1)^F$~\cite{Metlitski2017Majorana,AharonyOrthogonal2017}.
Table~\figtabref{tab:sm-background-responses} compares the two gapped branches, separating the bosonic response before and after the background counterterm.
For $r<0$, the full-rank Higgs configuration completely Higgses the dynamical $\mathrm{SO}(8)$, locking $b_v=A_{\mathrm{adj}}$.
By Eq.~\eqref{eq:sm-level-restriction} in Sec.~\ref{sec:sm-spin8-psu3}, this induces the response $-3\operatorname{CS}_{\mathfrak{su}(3)}[A]$, which the counterterm cancels, matching the trivial $m>0$ phase.
For $r>0$, the massive scalars leave $\mathrm{SO}(8)_{-1}$, an invertible spin TQFT whose only nonidentity line is the transparent fermion, as for $\mathrm{SO}(N)_{-1}$ generally~\cite{AharonyOrthogonal2017}.
Its response $8\operatorname{CS}_{\mathrm g}$ describes eight $p-\mathrm i p$ layers; adding the background counterterm gives $3\operatorname{CS}_{\mathfrak{su}(3)}[A]+8\operatorname{CS}_{\mathrm g}$, matching the $m<0$ phase without an additional gravitational counterterm.

\begin{table}[!htb]
\centering
\small
\renewcommand{\arraystretch}{1.3}
\begin{tabular}{@{}lccc@{}}
\toprule
\shortstack[l]{Paired\\branches}
& \shortstack{Bosonic response\\before counterterm}
& \shortstack{Full bosonic\\response}
& \shortstack{PV regularized\\fermionic response}\\
\midrule
$r<0$, $m>0$ & $-3\operatorname{CS}_{\mathfrak{su}(3)}$ & $0$ & $0$\\
$r>0$, $m<0$ & $8\operatorname{CS}_{\mathrm g}$
& $3\operatorname{CS}_{\mathfrak{su}(3)}+8\operatorname{CS}_{\mathrm g}$
& $3\operatorname{CS}_{\mathfrak{su}(3)}+8\operatorname{CS}_{\mathrm g}$\\
\bottomrule
\end{tabular}
\caption{Gapped-phase matching at fixed background $A$.
Entries are response three-forms, with the arguments $A$ and $\varpi$ suppressed; their integrals over $Y$, multiplied by $-\mathrm i$, give the topological Euclidean action.
The background counterterm adds $+3\operatorname{CS}_{\mathfrak{su}(3)}$ to the listed response in both branches.}
\label{tab:sm-background-responses}
\end{table}

Thus the level-three term already present in the CSH theory supplies precisely the counterterm required by the positive-PV free-fermion duality.
Promoting $A$ to the dynamical field $a$ and adding the same Yang--Mills term on both sides therefore gives the QCD description in Eq.~\eqref{eq:qcd3} without an additional bare Chern--Simons term.

\section{Emergent time reversal, spatial reflection, and the mass-operator dictionary}
\label{sec:sm-qcd-ct}

Time reversal can relate two ultraviolet Lagrangians that describe the same infrared physics~\cite{Seiberg2016DualityWeb,Metlitski2017Majorana}.
Here we formulate the proposed $\mathcal{CT}$ symmetry in terms of the QCD theory and its conjugate presentation, keeping track of the regulator and the mass deformation.

\paragraph*{The conjugate theory.}
Let $\widetilde{\mathcal L}_{\mathrm{QCD}}(\widetilde m)$ have the same local kinetic and Yang--Mills terms as Eq.~\eqref{eq:qcd3}, written in terms of an adjoint Majorana $\widetilde\chi$ and a $\mathrm{PSU}(3)$ connection $\widetilde a$, with mass term $+\widetilde m\operatorname{Tr}(\bar{\widetilde\chi}\widetilde\chi)$.
We keep the Clifford orientation fixed and use the conjugate PV prescription, $\widetilde M_{\mathrm{PV}}<0$, so that
\begin{equation}
 \widetilde Z_{\mathrm{adj}}(0)
 =|\operatorname{Pf}|\,e^{-\mathrm i\pi\eta_{\mathrm{adj}}[\widetilde a]/4},
 \qquad
 \widetilde I_{\mathrm{top}}(\widetilde m)
 =-\frac{1+\operatorname{sgn}\widetilde m}{2}
 \bigl(3I_{\mathrm{CS}}[\widetilde a]+8I_{\mathrm g}\bigr).
\label{eq:sm-qcd-conjugate-regulator}
\end{equation}
This is the $\mathrm{PSU}(3)_{-3/2}$ presentation.
The negative-mass branch has a trivial superconducting response, while the positive-mass branch gives $\mathrm{PSU}(3)_{-3}\boxtimes[p+\mathrm i p]^{\otimes8}$.
The ultraviolet $\mathcal{CT}$ operation reverses both the fermion mass and the regulator mass, taking $\mathcal L_{\mathrm{QCD}}(m)$ to $\widetilde{\mathcal L}_{\mathrm{QCD}}(-m)$.

\paragraph*{Matching the massive phases.}
For $m>0$ and $\widetilde m<0$, both presentations give $\mathrm{PSU}(3)$ Yang--Mills theory with a trivial superconductor.
For the other branch, level--rank duality relates $\mathrm{SU}(3)_3$ to the diagonal quotient $[\mathrm{SU}(3)_{-3}\boxtimes\mathrm{SU}(9)_1]/\mathbb Z_3$.
Gauging the remaining $\mathbb Z_3^{(1)}$ center symmetry gives~\cite{HsinSeiberg2016}
\begin{equation}
 \mathrm{PSU}(3)_3\simeq
 \mathrm{PSU}(3)_{-3}\boxtimes\frac{\mathrm{SU}(9)_1}{\mathbb Z_3}
 \simeq\mathrm{PSU}(3)_{-3}\boxtimes E_8.
\label{eq:sm-qcd-level-rank}
\end{equation}
The $\mathrm{SU}(9)_1$ theory has total quantum dimension $\mathcal D=3$ and chiral central charge $c_-=8$; gauging its bosonic $\mathbb Z_3^{(1)}$ removes all nontrivial anyons and leaves the invertible $E_8$ state.
As a spin phase, $E_8$ is equivalent to sixteen $p+\mathrm i p$ layers~\cite{Kitaev2006}.
Consequently,
\begin{equation}
 \mathrm{PSU}(3)_3\boxtimes[p-\mathrm i p]^{\otimes8}
 \simeq\mathrm{PSU}(3)_{-3}\boxtimes[p+\mathrm i p]^{\otimes8}.
\label{eq:qcd-ct-phase}
\end{equation}
The two sides are related by $\mathcal{CT}$ and have the same full topological order, including the invertible spin sector.
The ordered Yang--Mills branch also preserves the lattice time reversal: $\mathcal M\mapsto\mathcal M^\dagger$, together with antiunitarity, leaves its expectation value unchanged in each magnetic vacuum.

\paragraph*{Infrared duality and the tuning operator.}
Motivated by these phase matches, we propose the infrared duality
\begin{equation}
 \mathcal L_{\mathrm{QCD}}(m)
 \quad\longleftrightarrow\quad
 \widetilde{\mathcal L}_{\mathrm{QCD}}(-m).
\label{eq:sm-qcd-ct-duality}
\end{equation}
Near criticality, $m$ and $\widetilde m$ denote the corresponding renormalized tuning parameters; their positive relative normalization is absorbed into their definitions.
The tilded fields label the conjugate presentation, with the symmetry specified by the infrared equivalence of the two theories.
Assuming this duality, differentiating the paired mass deformations gives the operator correspondence
\begin{equation}
 \mathcal O_f=\operatorname{Tr}(\bar\chi\chi)
 \quad\longleftrightarrow\quad
 -\widetilde{\mathcal O}_f=-\operatorname{Tr}(\bar{\widetilde\chi}\widetilde\chi).
\label{eq:sm-qcd-mass-dictionary}
\end{equation}
The proposed emergent $\mathcal{CT}$ operation combines the ultraviolet transformation $\mathcal O_f\mapsto-\widetilde{\mathcal O}_f$ with the infrared identification in Eq.~\eqref{eq:sm-qcd-mass-dictionary}, which maps $-\widetilde{\mathcal O}_f$ back to $\mathcal O_f$.
The physical tuning operator is therefore $\mathcal{CT}$-even under this assumed infrared identification.
The Euclidean source terms are $+r\mathcal O_b$ and $+m\mathcal O_f$, where $\mathcal O_b=\operatorname{Tr}(\Phi^{\mathsf T}\Phi)$.
Since $m\propto-r$, their operator correspondence is
\begin{equation}
 \operatorname{Tr}(\Phi^{\mathsf T}\Phi)
 \quad\longleftrightarrow\quad
 -\operatorname{Tr}(\bar\chi\chi)
 \quad\longleftrightarrow\quad
 \operatorname{Tr}(\bar{\widetilde\chi}\widetilde\chi),
\label{eq:sm-boson-fermion-mass-dictionary}
\end{equation}
up to positive operator normalizations.

\paragraph*{Spatial reflection.}
The spatial part of $U_\rho$ is a proper rotation; that of $U_\tau$ is a mirror and cannot be absorbed into the proper Lorentz group.
Assuming the proposed $\mathcal{CT}$ symmetry, the CRT theorem also implies a unitary spatial reflection $\mathcal R$~\cite{Dumitrescu2026}.
Their combined action on the scalar monopole cancels time reversal and Hermitian conjugation, giving $\mathcal R:\mathcal M(t,x,y)\mapsto\mathcal M(t,-x,y)$ in real time.
Thus $U_\tau\leftrightarrow\mathcal{U}_\tau\mathcal R=\mathcal C\mathcal R$ sends $\mathcal M\mapsto\mathcal M^\dagger$ at the reflected point and $\mathcal{U}_\rho\mapsto\mathcal{U}_\rho^{-1}$, as required.
This $\mathcal R$ is implemented through the infrared duality; a bare spatial reflection reverses the ultraviolet Chern--Simons levels.
Consequently, lattice reflection and time reversal emerge together in the continuum description, while $\mathcal C$ remains an exact internal symmetry.
Both the uniform monopole condensate and the physical tuning operator are $\mathcal R$-even.
\end{document}